\documentstyle[12pt]{article}
\renewcommand{\baselinestretch}{1.5}

 1

\catcode`\@=11 
\def\lsim{\mathrel{\mathpalette\@versim<}}
\def\gsim{\mathrel{\mathpalette\@versim>}}
\def\@versim#1#2{\vcenter{\offinterlineskip
        \ialign{$\m@th#1\hfil##\hfil$\crcr#2\crcr\sim\crcr } }}
\makeatletter
\def\@seccntformat#1{\csname the#1\endcsname.\hskip 1em}
\makeatother

\begin{document}
\newcommand{\y}{{\it y}}
\newcommand{\as}{\alpha_s}
\newcommand{\A}{{\cal A}}
\newcommand{\B}{{\cal B}}
\newcommand{\Oa}{{\cal O}(\alpha_s)}
\newcommand{\Oaa}{{\cal O}(\alpha_s^2)}
\newcommand{\Oaaa}{{\cal O}(\alpha_s^3)}
\newcommand{\lam}{\Lambda_{\overline{MS}}}
\newcommand{\Rm}{{\it R-matching}}
\newcommand{\mRm}{{\it modified R-matching}}
\newcommand{\lnRm}{{\it lnR-matching}}
\newcommand{\mlnRm}{{\it modified lnR-matching}}
\newcommand{\aspi}{\tilde{\alpha}_s}
\newcommand{\asz}{\alpha_s(M_Z^2)}
 
\thispagestyle{empty}
\begin{flushright}
{\footnotesize\renewcommand{\baselinestretch}{.75}
  SLAC--PUB--6641\\
September 1994\\
(T/E)
}
\end{flushright}

\begin{center}
 {\Large \bf Measurement of $\asz$ from Hadronic Event Observables
                        at the $Z^0$ Resonance$^\star$}

\vspace {1.0cm}

 {\bf The SLD Collaboration$^*$}\\
Stanford Linear Accelerator Center \\
Stanford University, Stanford, CA~94309

\vspace{2.5cm}

\end{center} 

\normalsize


\begin{center}
{\bf ABSTRACT }
\end{center}

The strong coupling $\asz$ has been measured
using hadronic decays of $Z^0$ bosons collected by the SLD experiment at SLAC.
The data were compared with 
QCD predictions both at fixed order, ${\cal O}(\alpha_s^2)$,  
and including resummed analytic formulae based on the 
next-to-leading logarithm approximation.
In this comprehensive analysis we
 studied event shapes, jet rates, particle correlations,
and angular energy flow,
and checked the consistency between $\asz$ values extracted 
from these different measures.
Combining all results we obtain 
$\alpha_s(M_Z^2) = 0.1200 \pm 0.0025(\mbox{exp.}) \pm 0.0078(\mbox{theor.})$,
where the dominant uncertainty is from uncalculated
higher order contributions.
 
\vspace{0.5cm}
\begin{center}
{\it Submitted to Physical Review D.}
\end{center}
\vbox{
\uchyph=200
\brokenpenalty=200
\pretolerance=10000
\tolerance=2000
\nobreak
\penalty 5000
\hyphenpenalty=5000
\exhyphenpenalty=5000
\footnotesize\renewcommand{\baselinestretch}{1}\noindent
$^*$This work was supported by Department of Energy
  contracts:
  \hbox{DE-FG02-91ER40676 (BU),}
  DE-FG03-92ER40701 (CIT),
  DE-FG03-91ER40618 (UCSB),
  DE-FG03-92ER40689 (UCSC),
  DE-FG03-93ER40788 (CSU),
  DE-FG02-91ER40672 (Colorado),
  DE-FG02-91ER40677 (Illinois),
  DE-AC03-76SF00098 (LBL),
  DE-FG02-92ER40715 (Massachusetts),
  DE-AC02-76ER03069 (MIT),
  DE-FG06-85ER40224 (Oregon),
  DE-AC03-76SF00515 (SLAC),
  DE-FG05-91ER40627 (Tennessee),
  DE-AC02-76ER00881 (Wisconsin),
  DE-FG02-92ER40704 (Yale);
  National Science Foundation grants:
  PHY-91-13428 (UCSC),
  PHY-89-21320 (Columbia),
  PHY-92-04239 (Cincinnati),
  PHY-88-17930 (Rutgers),
  PHY-88-19316 (Vanderbilt),
  PHY-92-03212 (Washington);
  the UK Science and Engineering Research Council
  (Brunel and RAL);
  the Istituto Nazionale di Fisica Nucleare of Italy
  (Bologna, Ferrara, Frascati, Pisa, Padova, Perugia);
  and the Japan-US Cooperative Research Project on High Energy Physics
  (Nagoya, Tohoku).}
\newpage
 
\section{Introduction}

Achieving
precision tests of the Standard Model of elementary particle                   
interactions is one of the key aims of experimental high energy                
physics experiments. Some measurements in the electroweak sector 
have reached a precision of better than $1\%$ \cite{EW}. 
However,                     
measurements of strong                                                         
interactions, and hence tests of the theory of Quantum Chromodynamics          
(QCD) \cite{qcd}, have not yet achieved the same level of precision. 
This is largely due to the difficulty of performing QCD calculations, 
both at                
high order in perturbation theory and in the non-perturbative regime,       
where effects due to the hadronization process are important.               
QCD is a theory with only one free parameter, the strong coupling            
$\alpha_s$, which can be written in terms of a scale parameter $\lam$.
All tests of QCD can therefore be reduced to                    
a comparison of measurements of $\alpha_s$, either in different hard processes,
such as hadron-hadron collisions or e$^+$e$^-$ annihilations,
or at different energy scales $Q$.
In this paper we present                                                  
measurements of $\alpha_s$ in hadronic decays of $Z^0$                        
bosons produced by e$^+$e$^-$ annihilations at 
the SLAC Linear Collider and recorded          
in the SLC Large Detector.                                              
                                                                            
Complications arise in making accurate QCD predictions.               
In practice,
because of the large number of Feynman diagrams involved,
QCD calculations are only possible with                
present techniques to low order in perturbation theory. 
Perturbative calculations are performed within a particular                     
{\it renormalization scheme} \cite{scheme}, which also defines the strong               
coupling. Translation between different schemes is possible, without            
changing the final predictions, by appropriate redefinition of $\alpha_s$
and of the {\it renormalization scale} \cite{stan}. 
This leads to a {\it scheme-dependence}        
of $\alpha_s$, which can be alleviated in practice by choosing one
particular scheme as a standard and translating all $\alpha_s$ measurements
to it. 
The modified minimal subtraction scheme ($\overline{\mbox{MS}}$
 scheme) \cite{scheme} is presently used widely as this standard.                         
An additional complication is the 
truncation of the perturbative series at finite order, which             
yields a residual dependence on the renormalization scale,
often denoted by $\mu$ or equivalently by $f=\mu^2/Q^2$, 
which then becomes an arbitrary 
unphysical parameter.
 
In our previous studies of jet rates \cite{sld1} and energy-energy correlations       
\cite{sld2} it was shown that the dominant uncertainty in $\alpha_s(M_Z^2)$
measurements arises from this {\it renormalization scale ambiguity}.            
Given that infinite order perturbative QCD calculations would be                
independent of $\mu$, the scale uncertainty inherent in $\alpha_s$                    
measurements is a reflection of the neglected higher order terms.            
                                                                                
Distributions of observables in the process e$^+$e$^-$ $\rightarrow$ hadrons 
have been                
calculated exactly up to $\Oaa$ in QCD perturbation theory \cite{kn}.              
One expects {\it a priori} that the size of the uncalculated $\Oaaa$
and higher order terms will in general be different for each                    
observable, and hence that the scale dependence of the $\alpha_s$ values                  
measured using different observables will also be different. In order to        
make a realistic determination of $\alpha_s$ and its associated                    
theoretical uncertainty using $\Oaa$ calculations it is therefore              
advantageous to employ as many different observables as possible.               
Our previous measurements of $\alpha_s(M_Z^2)$ 
were based on extensive studies           
of jet rates \cite{jetrate} and energy-energy correlations and their asymmetry           
\cite{EEC}, using approximately 10,000 hadronic $Z^0$ decays collected by the          
SLD experiment in 1992. In this comprehensive analysis we have used the         
combined 1992 and 1993 data samples, comprising approximately 60,000 events, 
to make an improved determination of $\alpha_s(M_Z^2)$ 
using fifteen observables                  
presently calculated up to $\Oaa$ in perturbative QCD.                                             
                                                                                
In addition, for six of these fifteen 
observables, improved calculations can be formulated              
incorporating the resummation 
\cite{catani1,catani2,catani3,catani4,turnock,catani5}
of leading and next-to-leading               
logarithms matched to the $\Oaa$ results; these matched                        
calculations are expected {\it a priori} both                                   
to describe the data in a larger                                                
region of phase space than the fixed-order results, and to yield a              
reduced dependence of $\as$ on the renormalization scale. We have                
employed the matched calculations for all six observables to determine            
$\asz$, and have studied the uncertainties involved in the matching               
procedure. We have compared our results with our previous                       
measurements and with similar measurements from LEP.                            
                                                                                
We describe the detector and the event trigger and selection criteria              
applied to the data in Section 2. In Section 3 we define the                    
observables used to determine $\asz$ in this analysis. 
The QCD                
predictions are discussed in Section 4. The analysis of the data is            
described in Section 5, and a summary and conclusions are presented         
in Section~6.                                                                   

\section{Apparatus and Hadronic Event Selection}

The e$^+$e$^-$ annihilation events produced at the $Z^0$ resonance
by the SLAC Linear Collider (SLC)
have been recorded using the SLC Large Detector (SLD).
A general description of the SLD can be found elsewhere \cite{sld}.
Charged tracks are measured in the central drift 
chamber (CDC) 
and in the vertex detector (VXD) \cite{vxd}.
Momentum measurement is provided by a uniform axial magnetic field of 0.6 T.
Particle energies are measured in the Liquid Argon Calorimeter (LAC) 
\cite{lac},
which contains both electromagnetic and hadronic sections, 
and in the Warm Iron Calorimeter \cite{wic}.
 
Three triggers were used for hadronic events.
In the 1993 (1992) runs the first required
a total LAC
electromagnetic energy greater than 12 GeV (8 GeV);
the second required 
at least two well-separated tracks in the CDC; and the third required 
at least 4 GeV (8 GeV) in the LAC and one track in the CDC.
 A selection of hadronic events was then made by two independent methods,
one based on the topology of energy depositions in the calorimeters,
the other on the number and topology of charged tracks measured in the CDC.
 
The analysis presented here used the charged tracks measured 
in the CDC and VXD.
A set of cuts was applied to the data to select well-measured tracks 
and events well-contained within the detector acceptance.
The charged tracks were required to have
(i) a closest approach transverse to the beam axis within 5 cm,
and within 10 cm along the axis from the measured interaction point; 
(ii) a polar angle $\theta$ with respect to the beam axis within 
$\mid \cos \theta \mid < 0.80$; and
(iii)~a momentum transverse to the beam axis, $p_{\bot} > 0.15$ GeV/c.
Events were required to have
(i) a minimum of five such tracks;
(ii) a thrust axis \cite{thrust} direction 
within $\mid \cos \theta_T \mid < 0.71$; and
(iii)~a total visible energy $E_{vis}$ of at least 20~GeV,
which was calculated from the selected tracks assigned the charged pion mass.
From our 1992 and 1993 data samples 37,226 events passed these cuts.
The efficiency  for selecting hadronic events satisfying 
the $\mid \cos \theta_T \mid$ cut was estimated to be above $96\%$.
The background in the selected event sample was estimated to be $0.3\pm 0.1\%$,
dominated by $Z^0 \rightarrow \tau^+ \tau^-$ events. 
Distributions of single particle and event topology observables 
in the selected events were found to be well described by Monte Carlo models
of hadronic $Z^0$ decays \cite{jetset63,herwig}
combined with a simulation of the SLD.
 
\section{Definition of the Observables}

In this section we present the definitions of the quantities 
used in our measurement of $\asz$.
We used observables for which complete $\Oaa$ perturbative QCD 
calculations exist.
These include six event shapes, jet rates defined
by six schemes, two particle correlations, and an angular energy flow.
 
\subsection{Event Shapes}
 
Various inclusive observables have been proposed to describe the shapes of 
hadronic events in e$^+$e$^-$ annihilations. 
We considered those observables which are collinear
and infrared-safe, and which can hence be calculated in perturbative QCD.
 
Thrust $T$ is defined by \cite{thrust} 
\begin{eqnarray}
T = \mbox{max}\frac{\sum_i\mid \vec{p}_i \cdot \vec{n}_T \mid}
{\sum_i\mid \vec{p}_i \mid},
\label{thrust}
\end{eqnarray}
where $\vec{p}_i$ is the momentum vector of particle $i$,
and $\vec{n}_T$ is the thrust axis to be determined.
We define $\tau \equiv 1 - T$.
For back-to-back two-parton final states $\tau$ is zero,
while $0 \leq \tau \leq \frac{1}{3}$ for planar three-parton final states.
Spherical events have $\tau = \frac{1}{2}$.
An axis 
$\vec{n}_{maj}$ can be found to maximize the momentum sum transverse 
to $\vec{n}_T$.
Finally, an axis $\vec{n}_{min}$ is defined to be perpendicular to the two
axes $\vec{n}_T$ and $\vec{n}_{maj}$. The variables thrust-major $T_{maj}$ 
and thrust-minor $T_{min}$ 
are obtained by replacing $\vec{n}_T$ in Eq. (\ref{thrust})
by $\vec{n}_{maj}$ or $\vec{n}_{min}$, respectively. The oblateness $O$ is then
defined by \cite{oblate}
\begin{eqnarray}
O = T_{maj} - T_{min}.
\label{oblate}
\end{eqnarray}
The value of $O$ is zero for collinear or cylindrically symmetric final states,
and extends from zero to $1/{\sqrt 3}$ for three-parton final states.
 
The $C$-parameter is derived from the eigenvalues 
of the infrared-safe momentum tensor \cite{momten}:
\begin{eqnarray}
\theta_{\rho\sigma} = \frac{\sum_i p_i^\rho p_i^\sigma / \mid \vec{p}_i \mid}
                           {\sum_i \mid \vec{p}_i \mid},
\label{momtens}
\end{eqnarray}
where $p_i^\rho$ is the $\rho$-th component of the three momentum 
of particle $i$, and $i$ runs over all the final state particles.
The tensor $\theta_{\rho\sigma}$ is normalized 
to have unit trace, and the $C$-parameter is defined by:
\begin{eqnarray}
C = 3( \lambda_1 \lambda_2 + \lambda_2 \lambda_3 + \lambda_3 \lambda_1 ),
\label{cpar}
\end{eqnarray}
where $\lambda_i$ ($i = 1, 2, 3$) are the eigenvalues of the tensor 
$\theta_{\rho\sigma}$.
For back-to-back two-parton final states $C$ is zero,
while for planar three-parton final states
$0 \leq C \leq 2/3$. 
For spherical events $C = 1$.
 
Events can be divided into two hemispheres, $a$ and $b$, 
by a plane perpendicular to 
the thrust axis $\vec{n}_T$.
The heavy jet mass $M_H$ is then defined as \cite{jmass}
\begin{eqnarray}
M_H = \mbox{max}( M_a, M_b ),
\label{hevmas}
\end{eqnarray}
where $M_a$ and $M_b$ are the invariant masses of the two hemispheres.
Here we define the normalized quantity
\begin{eqnarray}
\rho \equiv \frac{M_H^2}{E_{vis}^2},
\label{rho}
\end{eqnarray}
where $E_{vis}$ is the total visible energy measured in hadronic events.
To first order in perturbative QCD, and for massless partons, 
the heavy jet mass and thrust are related by $\tau = \rho$ \cite{kn}.
 
Jet broadening measures have been proposed in Ref. \cite{jetb}.
In each hemisphere $a$, $b$:
\begin{eqnarray}
B_{a,b} = \frac{\sum_{i\in a,b} \mid \vec{p}_i \times \vec{n}_T \mid}
	       {2\sum_i \mid \vec{p}_i \mid}
\label{broad}
\end{eqnarray}
is calculated. The total jet broadening $B_T$ and wide jet broadening $B_W$
are defined by
\begin{eqnarray}
B_T = B_a + B_b~~~~\mbox{and}~~~~B_W = \mbox{max}(B_a,B_b),
\label{jetb}
\end{eqnarray}
respectively.
Both $B_T$ and $B_W$ are identically zero in two-parton final states and 
are sensitive to the transverse structure of jets.
To first order in perturbative QCD $B_T = B_W = \frac{1}{2}O$.
 
\subsection{Jet Rates}
 
Another useful method of classifying the structure of hadronic final states 
is in terms of jets.
Jets may be reconstructed using iterative clustering algorithms \cite{jetrate}
in which a measure $y_{ij}$, such as scaled invariant mass,
is calculated for all pairs of particles $i$ and $j$, and the pair with the 
smallest $y_{ij}$ is combined into a single particle.
This procedure is repeated until all pairs have $y_{ij}$ 
exceeding a value $y_{cut}$, and the jet multiplicity 
of the event is defined as the number of particles remaining.
The $n$-jet rate $R_n(y_{cut})$ is the fraction of events classified 
as $n$-jet,
and the differential 2-jet rate is defined as \cite{d2rate}
\begin{eqnarray}
D_2(y_{cut}) \equiv \frac{R_2(y_{cut}) - R_2(y_{cut} - \Delta y_{cut})}
                         {\Delta y_{cut} }.
\label{D2}
\end{eqnarray}
In contrast to $R_n$, each event contributes to $D_2$ at only one $y_{cut}$
value.
 
Several schemes have been proposed comprising different $y_{ij}$ definitions
and recombination procedures.
We have applied the E, E0, P, and P0 variations of the JADE algorithm \cite{jadea}
as well as the Durham (D) and Geneva (G) schemes \cite{jetrate}. 
The six definitions of the jet resolution parameter $y_{ij}$ and 
recombination procedure are given below.
 
In the E-scheme $y_{ij}$ is defined 
as the square of the invariant mass of the pair of
particles $i$ and $j$ scaled by the visible energy in the event,
\begin{eqnarray}
y_{ij} = \frac{(p_i + p_j)^2}{E_{vis}^2},
\label{e-resol}
\end{eqnarray}
with the recombination performed as
\begin{eqnarray}
p_k = p_i + p_j,
\label{e-recom}
\end{eqnarray}
where $p_i$ and $p_j$ are the four-momenta of the particles,
and pion masses are assumed in calculating particle energies.
Energy and momentum are explicitly conserved in this scheme.
 
The E0-, P-, and P0-schemes are variations of the E-scheme.
In the E0-scheme $y_{ij}$ is defined by 
Eq. (\ref{e-resol}), while the recombination is defined by 
\begin{eqnarray}
E_k       &=& E_i + E_j \\
\vec{p}_k &=& \frac{E_k}{\mid \vec{p}_i + \vec{p}_j \mid}
(\vec{p}_i + \vec{p}_j),
\label{e0-recom}
\end{eqnarray}
where $E_i$ and $E_j$ are the energies and  
$\vec{p}_i$ and $\vec{p}_j$ are the three-momenta of the particles.
The three-momentum $\vec{p}_k$ is rescaled so that particle $k$ 
has zero invariant mass.
This scheme does not conserve the total momentum sum of an event.
 
In the P-scheme
$y_{ij}$ is defined by Eq. (\ref{e-resol}) and the 
recombination is defined by
\begin{eqnarray}
\vec{p}_k &=& \vec{p}_i + \vec{p}_j \\
E_k       &=& \mid \vec{p}_k \mid.
\label{p-recom}
\end{eqnarray}
This scheme conserves the total momentum of an event,
but does not conserve the total energy.
 
The P0-scheme is similar to the P-scheme, but the total energy 
$E_{vis}$ in Eq. (\ref{e-resol}) is recalculated at each iteration according to
\begin{eqnarray}
E_{vis} = \sum_k E_k.
\label{p0-recom}
\end{eqnarray}
 
In the D-scheme 
\begin{eqnarray}
y_{ij} = \frac{2min(E_i^2,E_j^2)(1-\cos \theta_{ij})}{E_{vis}^2},
\label{d-resol}
\end{eqnarray}
where $\theta_{ij}$ is the angle between the pair of particles $i$ and $j$.
The recombination is defined by Eq. (\ref{e-recom}).
With the D-scheme a soft particle will be combined with another
soft particle, instead of being combined with a high-energy particle, 
only if the angle it makes with the other soft particle is smaller than 
the angle that it makes with the high-energy particle. 
 
The definition of $y_{ij}$ for the G-scheme is
\begin{eqnarray}
y_{ij} = \frac{8E_iE_j(1-\cos \theta_{ij})}{9(E_i+E_j)^2},
\label{g-resol}
\end{eqnarray}
and the recombination is defined by Eq. (\ref{e-recom}).
In this scheme soft particles are combined as in the D-scheme.
In addition, $y_{ij}$ depends only on the energy of the particles
to be combined, and not on the $E_{vis}$ of the event.
 
\subsection{Particle correlations}
 
Hadronic event observables can also be classified in terms of inclusive 
two-particle correlations.  
The energy-energy correlation ($EEC$) \cite{EEC} is the normalized 
energy-weighted cross section defined in terms of the angle $\chi_{ij}$ 
between two particles 
$i$ and $j$ in an event:
\begin{eqnarray}
EEC(\chi) \equiv \frac{1}{N_{events}\Delta \chi} 
\sum_{events} \int_{\chi -\frac{\Delta \chi}{2}}^{\chi +\frac{\Delta \chi}{2}}
\sum_{ij} \frac{E_iE_j}{E_{vis}^2}\delta(\chi' -\chi_{ij}) 
\mbox{d}\chi',
\label{EEC}
\end{eqnarray}
where $\chi$ is an opening angle to be studied for the correlations;
$\Delta\chi$ is the angular bin width; and
$E_i$ and $E_j$ are the energies of particles $i$ and $j$.
The angle $\chi$ is taken from $\chi=0^{\circ}~{\rm to}~180^{\circ}$.
The shape of the $EEC$ in the central region, $\chi \sim 90^{\circ}$,
is determined by hard gluon emission.
Hadronization contributions are expected to be large in the collinear and
back-to-back regions, $\chi \sim 0^{\circ}$ and $180^{\circ}$, respectively.
The asymmetry of the $EEC$ ($AEEC$) is defined as 
$AEEC(\chi) = EEC(180^{\circ}-\chi) - EEC(\chi)$.
 
\subsection{Angular Energy Flow}
 
Another procedure, related to the angle of particle emission, is to
integrate the energy within a conical shell 
of opening angle $\chi$ about the thrust axis.
The
{\it Jet Cone Energy Fraction} ($JCEF$) is defined \cite{jcef} as:
\begin{eqnarray}
JCEF(\chi) = \frac{1}{N_{events}\Delta\chi} \sum_{events}
\int_{\chi -\frac{\Delta\chi}{2}}^{\chi +\frac{\Delta\chi}{2}}
\sum_i \frac{E_i}{E_{vis}}\delta(\chi' - \chi_i)\mbox{d}\chi',
\label{djcef}
\end{eqnarray}
where
\begin{eqnarray}
\chi_i = \arccos \left( \frac{ \vec{p}_i \cdot
\vec{n}_T}{\mid\vec{p}_i\mid}\right)
\label{oangle}
\end{eqnarray}
is the opening angle between a particle and the thrust
axis vector, $\vec{n}_T$, whose direction is defined to point
from the heavy jet mass hemisphere to the light jet mass hemisphere,
and $0^{\circ} \leq \chi \leq 180^{\circ}$. 
Hard gluon emissions contribute to
the region corresponding to the heavy jet mass hemisphere, 
$90^{\circ} \leq \chi \leq 180^{\circ}$. 
 
\section{QCD Predictions}

The QCD predictions up to ${\cal O}(\alpha_s^2)$ for  all observables defined 
in Section 3 have the general form
\begin{eqnarray}
\frac{1}{\sigma_t}\cdot\frac{\mbox{d}\sigma(\y)}{\mbox{d}\y}
= A(\y)\widetilde{\alpha}_s + \left[B(\y)+A(\y)2\pi b_0\ln f\right]\aspi^2,
\label{kneq1}
\end{eqnarray}
where $\y$ is the observable in question;
$\sigma_t$ is the total hadronic cross section;
$\aspi = \as /2\pi$;
$f = \mu^2/s$; $b_0 = ({33-2n_f})/({12\pi})$; and  
$n_f$ is the number of active quark flavors; $n_f=5$ at 
$\sqrt{s}=M_Z$.
We have computed the coefficients $A(\y)$ and $B(\y)$ using 
the EVENT program, which was developed by Kunszt and Nason \cite{kn}.
It should be noted that a dependence on the QCD renormalization scale enters
explicitly in the second order term in Eq. (\ref{kneq1}).
 
It has been found recently \cite{catani1,catani2,catani3,catani4,
turnock,catani5}
that several observables, namely $\tau$, $\rho$, $B_T$, $B_W$, 
$D_2(D\mbox{-}\mbox{scheme})$, and $EEC$,
can be resummed, that is, leading and next-to-leading logarithmic terms 
can be calculated to all orders in $\as$ using an exponentiation technique.
This procedure 
is expected {\it a  priori} to yield formulae which are 
less dependent on the renormalization scale.
Using $L \equiv \ln (1/\y)$, the fraction $R(\y,\alpha_s)$ 
can then be written in the general form 
\begin{eqnarray}
R(\y,\alpha_s) &\equiv& 
\frac{1}{\sigma_t}\int_0^{\y}\frac{\mbox{d}\sigma}{\mbox{d}\y}\mbox{d}\y
= C(\alpha_s)\exp\{\Sigma(\alpha_s,L)\} + F(\y,\alpha_s),
\label{resum} 
\end{eqnarray}
where
\begin{eqnarray}
C(\alpha_s) &=& 1 + \sum_{n=1}^{\infty} C_n \aspi^n, \\
\label{c}
\Sigma(\alpha_s,L) &=& \sum_{n=1}^{\infty}\aspi^n\sum_{m=1}^{n+1}G_{nm}L^m,\\
\label{sigma1}
F(\y,\alpha_s) &=& \sum_{n=1}^{\infty} F_n(\y) \aspi^n.
\label{f}
\end{eqnarray}
The factor $\Sigma$ to be exponentiated can be written
\begin{eqnarray}
\Sigma(\alpha_s,L) = L\cdot f_{LL}(\alpha_sL) + f_{NLL}(\alpha_sL) + 
{\cal O}\left( \frac{1}{L} \cdot (\alpha_sL)^n \right),
\label{sigma2}
\end{eqnarray}
where $f_{LL}(\alpha_sL)$ and $f_{NLL}(\alpha_sL)$ are the leading 
and next-to-leading logarithms.
The functions $f_{LL}$ and $f_{NLL}$ depend only on the product  $\as L$
and are given in Refs.~\cite{catani1,catani2,catani3,catani4,turnock,catani5}. 
The resummed calculations are thus given by an approximate expression
for $R(\y,\alpha_s)$ in the form
\begin{eqnarray}
R^{resum}(\y,\alpha_s) = (1+C_1\aspi+C_2\aspi^2)\exp\{\Sigma^{resum}(
\alpha_s,L)\},
\label{nll1}
\end{eqnarray}
where
\begin{eqnarray}
\Sigma^{resum}(\alpha_s,L)=L\cdot f_{LL}(\alpha_sL)+f_{NLL}(
\alpha_sL).
\end{eqnarray}
 
Whereas the leading logarithmic ($L\cdot f_{LL}$) and 
next-to-leading logarithmic ($f_{NLL}$) terms in $\Sigma$ have been calculated,
the subleading terms in Eq.~(\ref{sigma2}) have not been 
completely computed.
However, some subleading terms included in $\Sigma$ (Eq. (25)),
as well as $C$ and $F$, are included in the $\Oaa$ calculation.
In order to make reliable predictions, 
including hard gluon emission, 
with the resummed calculations it is necessary to combine 
them with the second order calculations, taking overlapping terms into account.
This procedure is called {\it matching}, and four matching schemes have been 
proposed in the literature.
 
The $\Oaa$ QCD formula (Eq. (\ref{kneq1})) can also be cast into 
the integrated form
\begin{eqnarray}
R^{\Oaa}(\y,\alpha_s)  
 &=& 1 + \A(\y)\aspi + \B(\y)\aspi^2,
\label{kneq2}
\end{eqnarray}
where $\A(\y)$ and $\B(\y)$ are the cumulative forms of $A(\y)$ and $B(\y)$
in Eq. (\ref{kneq1}).
Taking the logarithm of the resummed formula (Eq. (\ref{nll1}))
and the ${\cal O}(\alpha_s^2)$ formula (Eq. (\ref{kneq2})), 
\begin{eqnarray}
\ln R^{resum}(\y,\alpha_s) &=& \Sigma^{resum}(\alpha_s,L) + C_1\aspi +
\left( C_2-\frac{C_1^2}{2} \right) \aspi^2+\Oaaa,
\label{nll2} 
\end{eqnarray}
and
\begin{eqnarray}
\ln R^{\Oaa}(\y,\alpha_s) &=& \A(\y)\aspi
+ \left( \B(\y) - \frac{\A^2(\y)}{2} \right) \aspi^2
+ \Oaaa.
\label{kneq3}
\end{eqnarray}
Adding Eq. (\ref{nll2}) and Eq. (\ref{kneq3}), and subtracting
the overlapping first and second order terms 
from Eq. (\ref{nll2}), yields  \cite{catani1,catani2}
\begin{eqnarray}
\ln R^{resum+\Oaa}(\y,\alpha_s) = \Sigma^{resum}(\alpha_s,L) -
\Sigma^{resum(1)}(\alpha_s,L) - \Sigma^{resum(2)}(\alpha_s,L) \nonumber \\
+ \A(\y)\aspi + \left( \B(\y) - \frac{\A^2(\y)}{2} \right) \aspi^2,
\label{lnR}
\end{eqnarray}
where
\begin{eqnarray}
\Sigma^{resum(1)}(\alpha_s,L) &=& G_{12}\aspi L^2 + G_{11}\aspi L \\
\Sigma^{resum(2)}(\alpha_s,L) &=& G_{23}\aspi^2 L^3 + G_{22}\aspi^2 L^2.
\end{eqnarray}
Finally, one can derive $R^{resum+\Oaa}(\y,\alpha_s)$ 
by taking the exponential of Eq. (\ref{lnR}).
This procedure is called \lnRm.
 
In an alternative approach, the overlapping terms 
$\Sigma^{resum(1)}(\alpha_s,L)$
and $\Sigma^{resum(2)}(\alpha_s,L)$ are subtracted from 
$\Sigma^{resum}(\alpha_s,L)$ in the form of an exponential. 
The exact formula up to ${\cal O}(\alpha_s^2)$ is then obtained as follows 
\cite{turnock,catani5}:
\begin{eqnarray}
R^{resum+\Oaa}(\y,\alpha_s) &=& (1+C_1\aspi+C_2\aspi^2)\left[ 
\exp\left\{ \Sigma^{resum}(\alpha_s,L) \right\} \right. \nonumber \\
& &\mbox{} - \left. \exp \left\{ \Sigma^{resum(1)}(\alpha_s,L) 
                               + \Sigma^{resum(2)}(\alpha_s,L) \right\} \right]
\nonumber \\
& &\mbox{} + 1 + \A(\y)\aspi + \B(\y)\aspi^2  \nonumber \\
&=&  (1+C_1\aspi+C_2\aspi^2)\exp\left\{\Sigma^{resum}(\alpha_s,L)\right\}
\nonumber \\
& &\mbox{} - (C_1\aspi + \Sigma^{resum(1)}(\alpha_s,L))
- \left[ C_2\aspi^2 + C_1\aspi\Sigma^{resum(1)}(\alpha_s,L) \right.
\nonumber \\
& &\mbox{} + \left. \frac{1}{2}\left\{\Sigma^{resum(1)}(\alpha_s,L)\right\}^2 
+ \Sigma^{resum(2)}(\alpha_s,L) \right] \nonumber \\
& &\mbox{} + \A(\y)\aspi + \B(\y)\aspi^2.
\label{R}
\end{eqnarray}
This is called {\it R-matching}, 
and differs from {\it lnR-matching} in that
the subleading term $G_{21}\aspi^2 L$ is not exponentiated.
In order to raise this procedure to the same level as the \lnRm \
scheme, Eq. (\ref{R}) may be modified by replacing 
$\Sigma^{resum}(\alpha_s,L)$ 
and
$\Sigma^{resum(2)}(\alpha_s,L)$ with 
$\Sigma(\alpha_s,L)$
and
$\Sigma^{(2)}(\alpha_s,L) = G_{23}\aspi^2 L^3+G_{22}\aspi^2 L^2
+G_{21}\aspi^2 L$, respectively. 
This procedure is called \mRm\footnote{
It has also been called {\it R-G$_{21}$-matching} \cite{RG21}, 
or {\it intermediate matching}
\cite{int}.} 
\cite{turnock}.
 
The predictions of these matching schemes have some troublesome features 
near the upper kinematic limit $\y_{max}$ 
because terms of third and higher order generated by 
the resummed calculations do not vanish at this limit.
This situation can be corrected by invoking a replacement of 
$L=\ln(1/\y)$ in Eq. (\ref{lnR}) 
with $L' = \ln(1/\y-1/\y_{max} + 1)$.
This procedure is called \mlnRm \ \cite{webber}.
We took the value of $\y_{max}$ to be 0.5 for $\tau$,
0.42 for $\rho$, 0.41 for $B_T$, 0.325 for $B_W$, and 0.33 for $D_2$(D).
 
Finally, in order to account for the renormalization scale dependence, 
$f_{NLL}(\alpha_sL)$ should be modified to 
$f_{NLL}(\alpha_sL)+(\alpha_sL)^2\frac{\mbox{d}f_{LL}(\alpha_sL)}
				      {\mbox{d}(\alpha_sL)}
b_0 \ln f$, and $\B(y)$ and $G_{22}$ should be modified to 
$\B(\y)+\A(\y)2\pi b_0\ln f$ and $G_{22}+G_{12}2\pi b_0\ln f$, respectively
\cite{kn,catani5}.
 
%
\section{Measurement of $\asz$}

\subsection{Data Analysis}

The fifteen observables defined in Section 3 were calculated from the
experimental data using
charged tracks in hadronic events selected according to the criteria
defined in Section 2. 
The experimental distributions $D^{data}_{SLD}(\y)$ were then
corrected for the effects of selection cuts,
detector acceptance, efficiency, and
resolution, for neutral particles, particle decays and interactions
within the detector, and for initial state photon radiation, using
bin-by-bin correction factors $C_D(\y)$:
\begin{eqnarray} 
C_D(\y)_i =  \frac{D^{MC}_{hadron}(\y)_i}{D^{MC}_{SLD}(\y)_i},
\label{cd}
\end{eqnarray} 
where $\y$ is the observable; $i$ is the bin index; 
$D^{MC}_{SLD}(\y)_i$ is the content of bin $i$ of the distribution 
obtained from reconstructed charged particles in Monte Carlo events 
after simulation of the detector; and
$D^{MC}_{hadron}(\y)_i$ is that from all generated particles with lifetimes
greater than $3 \times 10^{-10}$ s in Monte Carlo events 
with no SLD simulation and no initial state radiation.
The bin widths were chosen from the estimated experimental resolution 
so as to minimize bin-to-bin migration effects.
The $C_D(\y)$ were calculated using events generated with JETSET 6.3 
\cite{jetset63}
using parameter values tuned to hadronic e$^+$e$^-$ annihilation data 
\cite{tune}.
In addition, the multiplicity and momentum spectra of $B$ hadron decay
products were tuned to $\Upsilon_{4S}$ data \cite{cleo:argus}.
The {\it hadron level} distributions are then given by
\begin{eqnarray} 
D^{data}_{hadron}(\y)_i = C_D(\y)_i \cdot D^{data}_{SLD}(\y)_i.
\end{eqnarray}

Systematic effects were investigated using a variety of techniques.
The experimental systematic errors arising from uncertainties in
modeling the detector were estimated by varying the charged track and
event selection criteria over wide ranges, and by varying the tracking
efficiency and resolution in the detector simulation. 
In each case
the correction factors $C_D(\y)$, and hence the corrected data
distributions $D^{data}_{hadron}(\y)$, were rederived.
The data correction procedure was repeated by recalculating the
correction factors $C_D(\y)$ 
using events generated with HERWIG 5.5 \cite{herwig}.
In addition, a matrix correction procedure \cite{unfold} was employed,
in which 
migrations between all pairs of bins are accounted for individually.
The differences
between the data distributions corrected by the bin-by-bin and matrix
methods were
found to be much smaller than the statistical errors.

The hadron level
data are shown in Figs.~1--15 and
listed in Tables I--VII, together with statistical
and systematic errors; they may be compared with data from other
experiments that have applied corrections for detector effects.
The central values represent the data corrected
by the central values of the correction factors
$C_D(\y)$, which are shown in Figs.~16(c)--30(c).
For the $EEC$, $AEEC$, and $JCEF$, where there are
bin-to-bin correlations and multiple entries per event per bin, 
the statistical error in each bin was
estimated by taking the {\it rms} deviation of the contents of that bin 
over 50 Monte Carlo samples, each comprising
the same number of events as the data sample. The systematic errors
derive from the uncertainties on the correction factors shown in 
Figs. 16(c)--30(c).
Also shown in Figs. 1--15 are the
predictions of the JETSET 7.3 \cite{jetset73} and HERWIG 5.5 \cite{herwig} 
QCD + fragmentation event generators. Good agreement between the data and
model predictions is apparent in all cases.

Before they can be compared with the QCD predictions, the data must
be corrected for the effects of hadronization. 
The correction procedure is similar to that described above for 
the detector effects.
Bin-by-bin correction factors
\begin{eqnarray} 
C_H(\y)_i =  \frac{D^{MC}_{parton}(\y)_i}{D^{MC}_{hadron}(\y)_i},
\end{eqnarray} 
where $D^{MC}_{parton}(\y)_i$ is the
content of bin $i$ of the distribution obtained from Monte Carlo events
generated at the parton level, were calculated and applied to
the hadron level data distributions $D^{data}_{hadron}(\y)_i$ to obtain the
{\it parton level} corrected data:
\begin{eqnarray} 
D^{data}_{parton}(\y)_i = C_H(\y)_i \cdot D^{data}_{hadron}(\y)_i.
\end{eqnarray} 
The phenomenological hadronization models implemented in
JETSET 7.3 and HERWIG 5.5 were used to calculate the $C_H(\y)$.
In the case of JETSET the
$C_H(\y)$ were also recalculated for values of the parton virtuality
cutoff $Q_0$ \cite{jetset63,jetset73} in the range 0.5 to 2.0 GeV, 
and for reasonable variations of the parameters 
$\Lambda_{LL}$, $a$, and $\sigma_q$. 
The correction
factors $C_H(\y)$ are shown in Figs. 16(b)--30(b), where the bands
show the uncertainties due to model differences and parameter variations.
The parton level data are shown in Figs. 16(a)--30(a).
The data points
correspond to the central values of the hadronization correction factors,
and the errors shown are statistical and experimental systematic only;
the hadronization uncertainty will
be considered in the next sections which describe the fits to
determine $\asz$.
 
%
\subsection{Measurement of $\asz$ using ${\cal O}(\alpha_s^2)$ calculations}

We first determined $\asz$ by comparing the $\Oaa$ QCD calculations 
for each observable $\y$ with the corrected data at the parton level.
Each calculation was fitted to the
measured distribution $1/\sigma_t \cdot d\sigma/d\y$
by minimizing $\chi^2$ with respect to
variation of $\lam$.
In each $y$ bin $\chi^2$ was defined using the sum in quadrature of the
statistical and systematic errors.
Fits were performed at selected values of the scale $f$
and were restricted to the range in $y$ for which the $\Oaa$
calculation provides a good description of the corrected data.
 
The fit ranges in $\y$ were chosen to ensure that the parton level data
and the QCD calculations could be compared meaningfully. The range
for each observable
was determined according to the following requirements:
(1) the hadronization correction factors $C_H(\y)$ 
satisfied $0.6 < C_H(\y) < 1.4$;
(2) the systematic uncertainties on the
detector and hadronization correction factors, $\Delta C_D(\y)$ and
$\Delta C_H(\y)$ respectively, satisfied $\mid \Delta C_D(\y),
\Delta C_H(y) \mid < 0.3$; 
(3) three massless partons can contribute to the distribution 
at $\Oa$ in perturbative QCD; 
(4) the $\chi^2$ per degree of freedom, $\chi^2_{dof}$, 
for a fit at $f=1$ is 5.0 or less. 
Requirements (1) and (2) ensure that the corrected
data are well measured and that the hadronization corrections are
modeled reliably. Requirement (3) ensures that the kinematic regions
dominated by 4-parton production at $\Oaa$ are
excluded, as the calculation is effectively leading order, and hence
unreliable, in these regions. Requirement (4) is an empirical constraint
that ensures that the QCD calculation fits the data reasonably well;
this is most relevant to exclude the so-called `two-jet region' where multiple
emissions of soft or collinear gluons are important 
and are not included in the $\Oaa$ calculations,
a matter discussed further in Section 5.C.
These fit ranges are listed in Table VIII and are
shown in Figs. 16--30.
For illustration, fits to the distributions are shown in
Figs. 16(a)--30(a) for the case $f = 1$. 
The data are well described by $\Oaa$ QCD within the fit ranges. 
Fits were also performed {\it in the same ranges} for different choices 
of the renormalization scale $f$ such that $10^{-4} \leq f \leq 10^2$. 
In each case the fitted value of $\lam$ was translated \cite{pdg} to 
$\alpha_s(M_Z^2)$.
The value of $\alpha_s(M_Z^2)$ and the corresponding $\chi^2_{dof}$ for the
fit are shown as a function of the choice of $f$ in Figs. 31--33
for all observables.
 
Several features are common to the results from each observable: 
$\alpha_s(M_{Z}^2)$ depends strongly on $f$;
the fit quality is good over a wide range of $f$, 
typically $f \gsim 10^{-3}$,
and there is no strong preference for a particular scale 
for most of the observables;
at low $f$ the fit quality deteriorates rapidly, and neither $\asz$
nor its error can be interpreted meaningfully.
Similar features were reported in our earlier $\asz$ measurements from
jet rates \cite{sld1} and energy-energy correlations \cite{sld2}. 
For the oblateness the good fit region is
$f \gsim 10^{-1}$, which is much higher than for the other observables.
For $D_2$ calculated in the E-scheme the lowest $\chi^2_{dof}$ 
is found in the region around $f \sim 10^{-4}$,
which is much lower than for the other observables.
 
Figures 31--33 form a complete representation of the results of
the fits of $\Oaa$ QCD to our data.
It is useful, however, to quote a single
value of $\alpha_s(M_Z^2)$, together with its associated uncertainties,
determined from each observable. For this purpose we adopt the
following procedure, similar to that adopted in our
previous measurements \cite{sld1,sld2}.

For each observable  an $f$-range was defined such that $\chi^2_{dof} < 5.0$
and $f \leq 4.0$. The former requirement excludes the low $f$ regions where
the fit quality is poor, which has been shown \cite{pub6394} to be due to poor
convergence of the $\Oaa$ calculations. The latter requirement
corresponds to a reasonable physical limit $\mu \leq 2 \sqrt s$.
This range is arbitrary, but does ensure that the smallest $\alpha_s(M_Z^2)$
point (see Figs. 31(a)--33(a)) is considered for all variables 
except $B_T$. 
The extrema of $\asz$ values in this $f$-range were taken to define a symmetric
{\it renormalization scale uncertainty} about their average, which we defined 
as the central value.
The $f$-range,
central $\alpha_s(M_Z^2)$
value, and scale uncertainty are listed in Table VIII for each
observable.

For most observables the statistical error on $\alpha_s(M_Z^2)$
was defined by the change
in $\alpha_s(M_Z^2)$ corresponding to an increase in $\chi^2$ of 1.0 above
the lowest value within the $f$-range defined above (see 
Figs. 31(b)--33(b)). However, for the $EEC$, $AEEC$, and $JCEF$, 
where there are strong bin-to-bin correlations, 
the statistical error on $\alpha_s(M_Z^2)$ was
estimated by applying the same fitting procedure to ten sets of
Monte Carlo events, each comprising the same number of events as the
data sample, and taking the {\it rms} deviation over the ten samples. The
statistical error is less than 1\% of $\alpha_s(M_Z^2)$ for each
observable, and is listed in Table VIII.
 
For each observable the 
experimental systematic error on $\asz$ was estimated by 
changing the detector correction factor $C_D$ within
the systematic limits shown in Figs. 16(c)--30(c), and
by repeating 
the correction and fitting procedures to obtain $\lam$ and 
hence $\asz$ values.
The systematic error, calculated from the resulting
spread in $\asz$ values,
was found to be 1--3\% of $\asz$ for each observable
and is listed in Table VIII.
 
For each observable
the hadronization uncertainty on $\asz$ was 
estimated by changing the hadronization correction factor $C_H$
within the systematic limits shown in Figs. 16(b)--30(b),
and by repeating 
the correction and fitting procedures to obtain $\lam$ and 
hence $\asz$ values.
The hadronization uncertainty, calculated from the resulting
spread in $\asz$ values,
was found to be 0.4--6\% of $\asz$ for each observable
and is listed in Table VIII.
 
The central values of $\asz$ and the errors are summarized in Table IX.
For each observable the total experimental error is the sum in quadrature
of the statistical and experimental systematic errors,
and the total theoretical uncertainty is the sum in quadrature of the 
hadronization and scale uncertainties.
In all cases the theoretical uncertainty, which derives mainly from the scale
ambiguity, dominates.
This uncertainty, which arises from uncalculated higher order
terms in perturbation theory, varies from about 3\% of $\asz$ 
for the $AEEC$ to about 17\% of $\asz$ for $B_T$.
The $\asz$ values from the fifteen observables are consistent within these
theoretical uncertainties.
Since the same data were used to measure all observables,
and the observables are all highly correlated,
we combine these results using
an unweighted average to obtain
$$
\alpha_s(M_Z^2) = 0.1225 \pm 0.0026 (\mbox{exp.}) 
		            \pm 0.0109(\mbox{theor.}),
$$
where the experimental error is the sum in quadrature of 
the average statistical
($\pm0.0009$) and average experimental systematic ($\pm0.0024$) errors,
corresponding to the assumption that all are completely correlated.
The theoretical error is the sum in quadrature of the average hadronization
($\pm0.0024$) and average scale ($\pm0.0106$) uncertainties.
 
As a cross-check we combined the results by using weighted averages.
Weighting by experimental errors yields an average $\asz$ value different
from the above by $+0.0009$; 
weighting by the total errors yields an $\asz$ value different by $-0.0013$.
These differences are of the same order as the statistical error on a single
$\asz$ measurement and are hence negligible.
 
\subsection{Measurement of $\asz$ using resummed$+{\cal O}(\alpha_s^2)$ 
calculations}

We next determined $\asz$ by comparing the resummed$+\Oaa$ 
calculations with the corrected data at the parton level for those
observables for which
the resummed$+\Oaa$ calculations exist,
{\it i.e.} thrust ($\tau$), heavy jet mass ($\rho$), 
total ($B_T$) and wide ($B_W$) jet broadening measures, 
differential 2-jet rate ($D_2$) calculated in the D-scheme, 
and energy-energy correlations ($EEC$).
We considered all four matching schemes discussed in Section 4, 
namely, {\it lnR-}, {\it modified lnR-}, {\it R-}, and 
{\it modified R-matching}.
However, {\it modified R-matching} is not applicable to $D_2$ because 
the subleading term $G_{21}$\footnote{The value of $G_{21}$ cannot 
be estimated until a complete calculation of $G_{22}$ is 
available \cite{opal2}.} 
is not calculated in this case.
For the $EEC$ \lnRm \ and \mlnRm \ schemes cannot be
applied reliably \cite{opal2} and were not used.

The fit ranges were initially chosen to be the same as for the 
$\Oaa$ fits except for the $EEC$,
for which the fits were performed within the angular range 
$90^{\circ} \leq \chi \leq 154.8^{\circ}$, 
where the lower limit is the kinematic limit for the resummed$+\Oaa$ 
calculation.
For the fit to $D_2$ (D-scheme) we adopted a procedure \cite{sld1}
using the matched calculation for $0.03 \leq y_{cut} < 0.05$
and the ${\cal O}(\alpha_s^2)$ calculation for $0.05 \leq y_{cut} \leq 0.33$.
Fits to determine $\lam$, and hence $\asz$, were performed as described in
 the previous section. 
For illustration Figs. 16(a)--19(a), 26(a), and 28(a) show 
the results of the resummed$+{\cal O}(\alpha_s^2)$ QCD fits 
using the \mlnRm \ scheme
with the renormalization scale factor $f=1$. 
The data are well described by the QCD calculations within the fit ranges,
and also beyond the fit ranges into the so-called `two-jet region' or 
`Sudakov region' where the resummed contributions are 
large \cite{catani1,catani4}.   
This is discussed further at the end of this section.
Figures 34--37 show (a) $\alpha_s(M_Z^2)$
and (b) the corresponding $\chi^2_{dof}$, derived from fits at different
values of $f$, for the four matching schemes.
 
Several features should be noted from Figs. 34--37. 
For each matching scheme and each observable the
dependence of $\asz$ on $f$ (Figs. 34(a)--37(a))
is weaker than that from the $\Oaa$ fits
(Figs. 31(a)--33(a)); the range of $f$ for which the fit quality is
good (Figs. 34(b)--37(b)) is in all cases smaller than the
corresponding range from the $\Oaa$ fits (Figs. 31(b)--33(b)),
and some observables, most notably $B_T$ and $B_W$,
do display preferences for particular scales, typically in the range
$10^{-2} < f < 10$. 
However, using the \Rm \ scheme we found the fit qualities for 
$B_T$ and $B_W$ to be very poor for all scales. 
For a given observable, at any given $f$
the values of $\asz$ and $\chi^2_{dof}$ are typically similar for both of the
\lnRm \ schemes; however, the results from the two \Rm \
schemes are typically systematically different both between the two
schemes and with respect to the two \lnRm \ schemes. 
Since there is {\it a priori} no strong reason to reject individual 
matching schemes from consideration, 
it is necessary to consider an additional theoretical uncertainty
deriving from the matching ambiguity; this will be discussed below.
 
In order to quote a single $\asz$ value, and corresponding errors,
for each observable
we applied the same procedure as for the $\Oaa$ fits
to the results from each matching scheme.
Table X summarizes the $f$-ranges, central values of $\asz$, and scale
uncertainties.
The experimental and hadronization systematic uncertainties
were estimated by the methods described in Section 5.B 
and found to be similar to those from the $\Oaa$ analysis.
For each observable we then took the average $\asz$ value over all four
matching schemes. The maximum deviation of $\asz$ from the central
value was defined as the matching uncertainty, and was added in
quadrature with the hadronization and scale uncertainties to obtain
a total theoretical uncertainty for each observable.
The scale and matching uncertainties {\it both} derive from
uncalculated higher order perturbative contributions and are
therefore correlated, although to an unknown degree. 
The inclusion of both contributions in the total theoretical uncertainty 
therefore represents a conservative, though not unreasonable,
estimate of the effects of the higher
order contributions. 
The central $\asz$ value, 
total experimental error, defined as the sum in quadrature of the statistical
and experimental systematic errors, and the total theoretical
uncertainty are listed in Table XI.
 
Comparing the results in Tables IX and XI it is apparent that
the values of $\asz$ from the resummed$+\Oaa$ fits are lower
than those from the $\Oaa$ fits by about 3\% ($\tau$), 6\% ($\rho$), 
and 7\% ($B_T$ and $B_W$), but higher by about 4\% ($D_2$(D)) 
and 5\% ($EEC$).
In addition, for all
observables except $D_2$(D), the theoretical uncertainty is considerably
smaller for the resummed$+\Oaa$ case than for the $\Oaa$ case,
despite the extra matching uncertainty contribution to the former.
For $D_2$(D) the theoretical uncertainty is essentially the same for both
$\Oaa$ and resummed$+\Oaa$ cases, which may relate to the fact
that the resummation of next-to-leading logarithms of $y_{cut}$ to all
orders of $\as$ is not complete\footnote{A complete analytic expression has 
recently been obtained \cite{3jets}.} \cite{catani3}. 
In all cases, however, the theoretical uncertainty is larger 
than the experimental error.
 
Combining the resummed$+\Oaa$ results from all six observables 
using an unweighted average we obtain
$
\alpha_s(M_Z^2) = 0.1192 \pm 0.0025 (\mbox{exp.}) 
			    \pm 0.0070(\mbox{theor.}),
$
where the total experimental error is the sum in quadrature of 
the average statistical ($\pm0.0007$) and average experimental systematic 
($\pm0.0024$) errors,
and the total theoretical error is the sum in quadrature of 
the average hadronization($\pm0.0016$) and 
average scale and matching($\pm0.0065$) uncertainties.
As a cross-check we combined the results by using weighted averages.
Weighting by experimental errors yields an average $\asz$ value different
from the above by $-0.0011$; 
weighting by the total errors yields an $\asz$ value different by $-0.0015$.
These differences are of the same order as the statistical error on a single
$\asz$ measurement and are hence negligible.
 
It is interesting to compare the resummed$+\Oaa$ result with
the $\Oaa$ result. 
The final value quoted in Section 5.B is the average of the 
$\Oaa$ results over all 15 observables, 
whereas the value quoted above is the average of 
the resummed$+\Oaa$ results
over a subset of 6 observables. 
For the purposes of comparison we averaged the $\Oaa$ results 
for $\tau$, $\rho$, $B_T$, $B_W$, $D_2$(D), and $EEC$
to obtain $\asz = 0.1242 \pm 0.0026(\mbox{exp.}) \pm 0.0132(\mbox{theor.})$. 
For the {\it same set} of six observables, 
therefore, we find 
that the central $\asz$ values derived from $\Oaa$ and resummed$+\Oaa$
fits in the {\it same range} of each observable
are in agreement to within the (correlated) experimental errors, and  
that the theoretical uncertainty is significantly smaller when 
the resummed calculations are employed. 

From Figs. 16(a)--19(a), 26(a) and 28(a), it is clear that
the resummed$+{\cal O}(\alpha_s^2)$ calculations are more 
successful than the ${\cal O}(\alpha_s^2)$ calculations in describing
the two-jet (Sudakov) region.
This implies that multiple emissions of soft gluons,
which are taken into account in the resummed terms,
contribute significantly to this region.
Therefore, for each observable
we extended the fit range into the two-jet region and extracted $\asz$ as a 
function of the renormalization scale factor $f$.
Requirements (1)--(3) (Section 5.B) were applied.
In addition,  
for $D_2$(D) we required the 5-jet production rate $R_5$
to be less than 1\%;
for the $EEC$ the upper limit of the fit range was extended to 
$\chi = 162^{\circ}$ by applying the empirical criterion $\chi^2_{dof} < 5$. 
The fit ranges are listed in Table XII.

The same procedure as above was applied to define a range of
renormalization scale factor $f$ over which to calculate a central
$\asz$ value and scale uncertainty for each observable; the $f$-range,
central $\asz$ value, and scale uncertainty are listed in Table~XII
separately for fits using each of the four matching schemes.
Good fits with $\chi^2_{dof} < 5$ could not be obtained using the \Rm \
scheme for $\tau$, $B_T$, $B_W$, and $D_2$(D) for any extension of the
fit range beyond that used for the $\Oaa$ fits.
By comparing Tables X and XII it can be seen that the maximum change in
$\asz$ when the fit range is extended into the two-jet region is $-0.0026$
for $\tau$ (\lnRm), $-0.0038$ for $\rho$ (\Rm), $-0.0009$ for
$B_T$ (\mlnRm), $-0.0006$ for $B_W$ (\mlnRm),
$-0.0045$ for $D_2$(D) (\mlnRm), and $-0.0006$ for
the $EEC$ (\Rm). 
These shifts are smaller than, or comparable with,
the experimental errors, and are much smaller than the theoretical
uncertainties. 
 
For each observable the average $\asz$ value over all four matching
schemes, and the matching uncertainty, were calculated as before.
The central $\asz$ value, the total experimental error,
and the total theoretical uncertainty, defined as before, are listed in
Table XIII.
Averaging over the six observables, as above, then yields
$$ 
\asz = 0.1181 \pm 0.0024(\mbox{exp.}) \pm 0.0057(\mbox{theor.}),
$$ 
which is in good agreement with the above average of results from the 
restricted fit ranges.
%
 
\section{Conclusions}

We have measured the strong coupling
$\alpha_s(M_Z^2)$ by analyses of fifteen different
observables that describe the hadronic final states of 
about 60,000 $Z^0$ decays recorded by the SLD experiment.
The observables comprise                                                      
six event shapes ($\tau$, $\rho$, $B_T$, $B_W$, $O$, and $C$),                 
differential 2-jet rates ($D_2$) defined by six different
jet resolution/recombination schemes (E, E0, P, P0, D, and G), 
energy-energy correlations ($EEC$) and their asymmetry ($AEEC$),
and the jet cone energy fraction ($JCEF$).                                     
The quantity $JCEF$ has been measured for the first time.                
Our measured distributions of these observables are reproduced 
by the JETSET and HERWIG Monte Carlo simulations of hadronic $Z^0$ decays.
The coupling was determined by fitting perturbative QCD calculations
to the data corrected to the parton level.                      
Perturbative QCD calculations complete to $\Oaa$ were used 
for all 15 observables. 
In addition, recently-performed resummed calculations 
were matched to the $\Oaa$ calculations      
using four matching schemes and applied to the six observables 
for which the resummed calculations are available. 
 
We find that the $\Oaa$ calculations are able to describe the data 
in the hard 3-jet region of all 15 observables for a wide range 
of the QCD renormalization scale factor $f$.         
The fitted $\asz$ value depends strongly both on the choice of $f$, 
which limits the precision of the $\asz$ measurement
from each observable, and on the choice of observable. 
The $AEEC$ shows the smallest renormalization scale 
uncertainty of about 3\%, which is just larger than the experimental error.
The $\asz$ values determined from jet rates and energy-energy correlations are
consistent with our previous measurements \cite{sld1,sld2}
within experimental errors.                       
The $\asz$ values from the various observables are consistent with 
each other only within the scale uncertainties.  
The large scale uncertainties and systematically different $\asz$ values 
determined from different observables imply that the uncalculated $\Oaaa$
perturbative QCD contributions are significant and cannot be ignored
if $\asz$ is to be determined with a precision of better than 10\%.
 
The resummed$+\Oaa$ calculations yield a reduced 
renormalization scale dependence of $\asz$, 
and fit a wider kinematic region, 
including the two-jet or Sudakov region,
and give similar fitted values of $\asz$ to the $\Oaa$ case. 
However, the different matching schemes give different 
$\asz$ values, which reflects a residual uncertainty 
in the inclusion of terms in the resummed$+\Oaa$ calculations. 
For all observables except $D_2$(D) the theoretical uncertainty is smaller 
than in the $\Oaa$ case, 
but still dominates the uncertainty in 
the measurement of $\asz$. 
Again, the $\asz$ values derived from jet rates and energy-energy correlations
are consistent with 
our previous measurements \cite{sld1,sld2} within experimental errors, 
and the values determined from the six observables are consistent 
within theoretical uncertainties.                                      
 
Figure 38 summarizes the measured $\asz$ values from all 
fifteen observables using $\Oaa$ calculations, and from the six observables 
using resummed$+\Oaa$ calculations in the extended kinematic region.
Since the same data were used to measure all observables, 
and the observables are highly correlated,
we combined the results by taking unweighted averages of the $\asz$ values
and experimental and theoretical errors, obtaining 
\begin{eqnarray}
\alpha_s(M_Z^2) &=& 0.1225 \pm 0.0026({\rm exp.})
\pm 0.0109({\rm theor.})~~~~~~\Oaa \nonumber \\
\alpha_s(M_Z^2) &=& 0.1181 \pm 0.0024({\rm exp.}) 
\pm 0.0057({\rm theor.})~~~~~~\mbox{resummed}+\Oaa , \nonumber
\end{eqnarray} 
where in both cases the theoretical uncertainty 
is dominated by the lack of knowledge
of higher order terms in the QCD calculations.  
Our estimate of the theoretical uncertainty is larger than that quoted by 
some of the LEP experiments because we have considered more observables 
and wider variations of the renormalization scale, 
and have taken unweighted averages. 
These average values are shown in Fig. 38;
they are consistent 
with measurements from other e$^+$e$^-$
experiments at the $Z^0$ resonance \cite{d2rate,RG21,int,opal2,LEP}
and from lower energy e$^+$e$^-$ and deep inelastic scattering
experiments~\cite{expsum}.
 
One expects {\it a priori} the $\asz$ value determined from a 
resummed$+\Oaa$ fit to be more reliable than that from an $\Oaa$ fit.
However, the former is only available for six of the fifteen observables.
In order to quote a final result, therefore, we took the unweighted 
average of the $\asz$ values and uncertainties 
over the combined set of six resummed$+\Oaa$ results and nine $\Oaa$
results for which there is no corresponding resummed$+\Oaa$ result.
This yields a final average of 
$$
\asz = 0.1200 \pm 0.0025 (\mbox{exp.}) \pm 0.0078 (\mbox{theor.}), 
$$                                                                
also shown in Fig. 38, corresponding to $\lam = 253^{+130}_{-~96}$ MeV.
 
 
\section*{Acknowledgements} 

We thank the personnel of the SLAC accelerator department 
and the technical staffs of our collaborating institutions for their
efforts which resulted in the successful operation of the SLC and the SLD.
We also thank S.J. Brodsky, S.D. Ellis, K. Kato, P. Nason, and D. Ward
for helpful comments and suggestions relating to this analysis.


\section*{List of Authors}
{
%
%
%
  \def\iADEL{$^{(1)}$}
  \def\iBOL{$^{(2)}$}
  \def\iBU{$^{(3)}$}
  \def\iBRUN{$^{(4)}$}
  \def\iCIT{$^{(5)}$}
  \def\iUCSB{$^{(6)}$}
  \def\iUCSC{$^{(7)}$}
  \def\iCIN{$^{(8)}$}
  \def\iCSU{$^{(9)}$}
  \def\iCOLO{$^{(10)}$}
  \def\iCOL{$^{(11)}$}
  \def\iFER{$^{(12)}$}
  \def\iFRA{$^{(13)}$}
  \def\iILL{$^{(14)}$}
  \def\iLBL{$^{(15)}$}
  \def\iMIT{$^{(16)}$}
  \def\iMASS{$^{(17)}$}
  \def\iMISS{$^{(18)}$}
  \def\iNAG{$^{(19)}$}
  \def\iOREG{$^{(20)}$}
  \def\iPAD{$^{(21)}$}
  \def\iPERU{$^{(22)}$}
  \def\iPISA{$^{(23)}$}
  \def\iRUT{$^{(24)}$}
  \def\iRAL{$^{(25)}$}
  \def\iSOGANG{$^{(26)}$}
  \def\iSLAC{$^{(27)}$}
  \def\iTENN{$^{(28)}$}
  \def\iTOH{$^{(29)}$}
  \def\iVAND{$^{(30)}$}
  \def\iWASH{$^{(31)}$}
  \def\iWISC{$^{(32)}$}
  \def\iYALE{$^{(33)}$}
  \def\dead{$^{\dag}$}
  \def\andgen{$^{(a)}$}
  \def\andper{$^{(b)}$}
%
%
\begin{flushleft}
{$^*$
\mbox{K. Abe                 \unskip,\iTOH}
\mbox{I. Abt                 \unskip,\iILL}
\mbox{C.J. Ahn               \unskip,\iSOGANG}
\mbox{T. Akagi               \unskip,\iSLAC}
\mbox{W.W. Ash               \unskip,\iSLAC$^\dagger$}
\mbox{D. Aston               \unskip,\iSLAC}
\mbox{N. Bacchetta           \unskip,\iPAD}
\mbox{K.G. Baird             \unskip,\iRUT}
\mbox{C. Baltay              \unskip,\iYALE}
\mbox{H.R. Band              \unskip,\iWISC
% ...
}}
\end{flushleft}
%
\begin{center}
{(The SLD Collaboration)}   
\it
  \vskip \baselineskip                   
%
%
%
  \iADEL
     Adelphi University,
     Garden City, New York 11530 \break
  \iBOL
     INFN Sezione di Bologna,
     I-40126 Bologna, Italy \break
  \iBU
     Boston University,
     Boston, Massachusetts 02215 \break
  \iBRUN
     Brunel University,
     Uxbridge, Middlesex UB8 3PH, United Kingdom \break
  \iCIT
     California Institute of Technology,
     Pasadena, California 91125 \break
  \iUCSB
     University of California at Santa Barbara,
     Santa Barbara, California 93106 \break
  \iUCSC
     University of California at Santa Cruz,
     Santa Cruz, California 95064 \break
  \iCIN
     University of Cincinnati,
     Cincinnati, Ohio 45221 \break
  \iCSU
     Colorado State University,
     Fort Collins, Colorado 80523 \break
  \iCOLO
     University of Colorado,
     Boulder, Colorado 80309 \break
  \iCOL
     Columbia University,
     New York, New York 10027 \break
  \iFER
     INFN Sezione di Ferrara and Universit\`a di Ferrara,
     I-44100 Ferrara, Italy \break
  \iFRA
     INFN  Lab. Nazionali di Frascati,
     I-00044 Frascati, Italy \break
  \iILL
     University of Illinois,
     Urbana, Illinois 61801 \break
  \iLBL
     Lawrence Berkeley Laboratory, University of California,\break
     Berkeley, California 94720 \break
  \iMIT
     Massachusetts Institute of Technology,
     Cambridge, Massachusetts 02139 \break
  \iMASS
     University of Massachusetts,
     Amherst, Massachusetts 01003 \break
  \iMISS
     University of Mississippi,
     University, Mississippi  38677 \break
  \iNAG
     Nagoya University,
     Chikusa-ku, Nagoya 464 Japan  \break
  \iOREG
     University of Oregon,
     Eugene, Oregon 97403 \break
  \iPAD
     INFN Sezione di Padova and Universit\`a di Padova,
     I-35100 Padova, Italy \break
  \iPERU
     INFN Sezione di Perugia and Universit\`a di Perugia,
     I-06100 Perugia, Italy \break
  \iPISA
     INFN Sezione di Pisa and Universit\`a di Pisa,
     I-56100 Pisa, Italy \break
  \iRUT
     Rutgers University,
     Piscataway, New Jersey 08855 \break
  \iRAL
     Rutherford Appleton Laboratory,\break
     Chilton, Didcot, Oxon OX11 0QX United Kingdom \break
 \iSOGANG
     Sogang University, Seoul Korea \break
  \iSLAC
     Stanford Linear Accelerator Center, Stanford University,\break
     Stanford, California 94309 \break
  \iTENN
     University of Tennessee,
     Knoxville, Tennessee 37996 \break
  \iTOH
     Tohoku University,
     Sendai 980 Japan \break
  \iVAND
     Vanderbilt University,
     Nashville, Tennessee 37235 \break
  \iWASH
     University of Washington,
     Seattle, Washington 98195 \break
  \iWISC
     University of Wisconsin,
     Madison, Wisconsin 53706 \break
  \iYALE
     Yale University,
     New Haven, Connecticut 06511 \break
  \dead
     Deceased \break
  \andgen
     Also at the Universit\`a di Genova \break
  \andper
     Also at the Universit\`a di Perugia \break
\rm
%
\end{center}}
 
\begin{table}[t]
 
\begin{center}
\footnotesize
\renewcommand{\arraystretch}{0.8}
\begin{tabular}{|l|c||l|c|}  \hline
$\tau$ & $\frac{1}{\sigma_t}\frac{d\sigma}{d\tau}\pm (stat.)
\pm (exp.~sys.)$ &
$\rho$ & $\frac{1}{\sigma_t}\frac{d\sigma}{d\rho}\pm (stat.)
\pm (exp.~sys.)$ \\
\hline
$0.0~-0.02$    	 & $~~7.01 \pm 0.10~~ \pm 0.50~~$ &
$0.0~-0.02$	 & $~10.53 \pm 0.12~~ \pm 0.41~~$ 
\\
$0.02-0.04$	 & $~16.10 \pm 0.15~~ \pm 0.15~~$ &
$0.02-0.04$	 & $~17.38 \pm 0.15~~ \pm 0.14~~$ 
\\
$0.04-0.06$	 & $~~8.67 \pm 0.11~~ \pm 0.05~~$ &
$0.04-0.08$	 & $~~6.21 \pm 0.07~~ \pm 0.16~~$ 
\\
$0.06-0.08$	 & $~~5.08 \pm 0.08~~ \pm 0.16~~$ &
$0.08-0.12$	 & $~~2.39 \pm 0.04~~ \pm 0.09~~$ 
\\
$0.08-0.12$	 & $~~2.91 \pm 0.04~~ \pm 0.06~~$ &
$0.12-0.18$	 & $~~1.08 \pm 0.02~~ \pm 0.04~~$ 
\\
$0.12-0.16$	 & $~~1.57 \pm 0.03~~ \pm 0.05~~$ &
$0.18-0.24$	 & $~0.404 \pm 0.014~ \pm 0.021~$ 
\\
$0.16-0.20$	 & $~0.917 \pm 0.025~ \pm 0.028~$ &
$0.24-0.32$	 & $~0.102 \pm 0.006~ \pm 0.010~$ 
\\
$0.20-0.26$	 & $~0.495 \pm 0.015~ \pm 0.025~$ &
$0.32-0.40$	 & $0.0047 \pm 0.0013 \pm 0.0008$ 
\\
$0.26-0.32$	 & $~0.227 \pm 0.010~ \pm 0.016~$ &
	 	 & 
\\
$0.32-0.38$	 & $~0.061 \pm 0.005~ \pm 0.006~$ &
         	 & 
\\
$0.38-0.44$	 & $~0.003 \pm 0.001~ \pm 0.003~$ &
	  	 &
\\ \hline
\end{tabular}
\end{center}
\normalsize
 
Table I.
Distributions of $\tau$ and $\rho$  (see text).
The data were corrected for detector effects and 
for initial state photon radiation.
The first error is statistical, and 
the second represents the experimental systematic uncertainty.
 
\end{table}
\begin{table}[b]
 
\begin{center}
\footnotesize
\renewcommand{\arraystretch}{0.8}
\begin{tabular}{|l|c||l|c|}  \hline
$B_T$ & $\frac{1}{\sigma_t}\frac{d\sigma}{dB_T}\pm (stat.)\pm (exp.~sys.)$ &
$B_W$ & $\frac{1}{\sigma_t}\frac{d\sigma}{dB_W}\pm (stat.)\pm (exp.~sys.)$ \\
\hline
$0.0~-0.02$ 	 & $0.018 \pm 0.005 \pm 0.007$ &
$0.0~-0.02$      & $0.570 \pm 0.028 \pm 0.213$ 
\\
$0.02-0.04$	 & $~1.36 \pm 0.04~ \pm 0.18~$ &
$0.02-0.04$	 & $13.86 \pm 0.14~ \pm 0.45~$ 
\\
$0.04-0.06$	 & $~8.81 \pm 0.11~ \pm 0.32~$ &
$0.04-0.06$	 & $11.71 \pm 0.13~ \pm 0.20~$ 
\\
$0.06-0.08$	 & $10.64 \pm 0.12~ \pm 0.16~$ &
$0.06-0.08$	 & $~7.38 \pm 0.10~ \pm 0.11~$ 
\\
$0.08-0.12$	 & $~6.52 \pm 0.07~ \pm 0.10~$ &
$0.08-0.12$	 & $~4.29 \pm 0.05~ \pm 0.08~$ 
\\
$0.12-0.16$	 & $~3.65 \pm 0.05~ \pm 0.04~$ &
$0.12-0.16$	 & $2.185 \pm 0.038 \pm 0.128$ 
\\
$0.16-0.20$	 & $~2.10 \pm 0.04~ \pm 0.06~$ &
$0.16-0.20$	 & $~1.12 \pm 0.028 \pm 0.061$ 
\\
$0.20-0.26$	 & $~1.12 \pm 0.02~ \pm 0.03~$ &
$0.20-0.26$	 & $0.403 \pm 0.014 \pm 0.025$ 
\\
$0.26-0.32$	 & $0.384 \pm 0.013 \pm 0.023$ &
$0.26-0.32$	 & $0.403 \pm 0.014 \pm 0.025$ 
\\
$0.32-0.38$	 & $0.050 \pm 0.005 \pm 0.011$ &
$0.32-0.38$	 & $0.030 \pm 0.004 \pm 0.005$ 
\\ \hline
\end{tabular}
\end{center}
\normalsize
 
Table II.
Distributions of $B_T$ and $B_W$ (see text).
The data were corrected for detector effects and 
for initial state photon radiation.
The first error is statistical, and 
the second represents the experimental systematic uncertainty.
 
\end{table}
 
\begin{table}[t]
 
\begin{center}
\footnotesize
\renewcommand{\arraystretch}{0.8}
\begin{tabular}{|l|c||l|c|}  \hline
$O$ & $\frac{1}{\sigma_t}\frac{d\sigma}{dO}\pm (stat.)\pm (exp.~sys.)$ &
$C$ & $\frac{1}{\sigma_t}\frac{d\sigma}{dC}\pm (stat.)\pm (exp.~sys.)$ \\
\hline
$0.0~-0.02$ & $~9.07 \pm 0.11~ \pm 0.19~$ &
$0.0~-0.04$ & $0.166 \pm 0.011 \pm 0.015$
\\
$0.02-0.04$ & $11.28 \pm 0.12~ \pm 0.20~$ &
$0.04-0.08$ & $~1.76 \pm 0.03~ \pm 0.04~$
\\
$0.04-0.08$ & $~5.98 \pm 0.06~ \pm 0.07~$ &
$0.08-0.12$ & $~4.01 \pm 0.05~ \pm 0.09~$ 
\\
$0.08-0.12$ & $~3.16 \pm 0.05~ \pm 0.06~$ &
$0.12-0.18$ & $~3.57 \pm 0.04~ \pm 0.10~$ 
\\
$0.12-0.18$ & $~1.77 \pm 0.03~ \pm 0.03~$ &
$0.18-0.24$ & $~2.30 \pm 0.03~ \pm 0.02~$ 
\\
$0.18-0.24$ & $0.935 \pm 0.021 \pm 0.028$ &
$0.24-0.32$ & $~1.54 \pm 0.02~ \pm 0.016$ 
\\
$0.24-0.32$ & $0.523 \pm 0.013 \pm 0.013$ &
$0.32-0.40$ & $~1.07 \pm 0.02~ \pm 0.03~$
\\
$0.32-0.40$ & $0.223 \pm 0.009 \pm 0.010$ &
$0.40-0.52$ & $0.718 \pm 0.013 \pm 0.024$
\\
$0.40-0.50$ & $0.052 \pm 0.004 \pm 0.003$ &
$0.52-0.64$ & $0.491 \pm 0.011 \pm 0.013$
\\
& &
$0.64-0.76$ & $0.311 \pm 0.008 \pm 0.022$
\\
& &
$0.76-0.88$ & $0.146 \pm 0.006 \pm 0.012$ 
\\
& &
$0.88-1.0~$ & $0.012 \pm 0.002 \pm 0.001$
\\ \hline
\end{tabular}
\end{center}
\normalsize
 
Table III.
Distributions of $O$ and $C$ (see text).
The data were corrected for detector effects and 
for initial state photon radiation.
The first error is statistical, and 
the second represents the experimental systematic uncertainty.
 
\end{table}
\begin{table}[t]
 
\begin{center}
\footnotesize
\renewcommand{\arraystretch}{0.8}
\begin{tabular}{|c|c|c|c|}  \hline
        & E-scheme & E0-scheme & P-scheme~ \\
\raisebox{1.5ex}{$y_{cut}$} & $D_2(y_{cut}) \pm (stat.) \pm (exp.~sys.)$ &
            $D_2(y_{cut}) \pm (stat.) \pm (exp.~sys.)$ &
            $D_2(y_{cut}) \pm (stat.) \pm (exp.~sys.)$ \\
\hline
0.005   & $0.669 \pm 0.060 \pm 0.080$ & 
          $28.95 \pm 0.39~ \pm 1.44~$ & 
	  $41.80 \pm 0.47~ \pm 2.43~$
\\
0.010   & $~2.60 \pm 0.12~ \pm 0.12~$ & 
          $25.25 \pm 0.37~ \pm 0.50~$ &  
	  $31.06 \pm 0.41~ \pm 0.63~$
\\
0.015   & $~7.07 \pm 0.20~ \pm 0.27~$ &
          $19.93 \pm 0.33~ \pm 0.53~$ &
	  $21.24 \pm 0.34~ \pm 0.28~$
\\
0.02    & $10.48 \pm 0.24~ \pm 0.66~$ &  
          $15.85 \pm 0.29~ \pm 1.04~$ &
	  $14.96 \pm 0.28~ \pm 0.54~$
\\
0.03    & $12.28 \pm 0.18~ \pm 0.39~$ &
          $11.66 \pm 0.18~ \pm 0.15~$ &
	  $10.82 \pm 0.17~ \pm 0.37~$
\\
0.05    & $10.89 \pm 0.12~ \pm 0.34~$ &
          $~7.01 \pm 0.10~ \pm 0.19~$ &
	  $~6.35 \pm 0.09~ \pm 0.23~$
\\
0.08    & $~7.22 \pm 0.08~ \pm 0.22~$ &
          $~3.85 \pm 0.06~ \pm 0.05~$ &
	  $~3.16 \pm 0.05~ \pm 0.09~$
\\
0.12    & $~3.81 \pm 0.05~ \pm 0.11~$ &  
          $~2.02 \pm 0.04~ \pm 0.07~$ &
	  $~1.61 \pm 0.03~ \pm 0.08~$
\\
0.17    & $~1.97 \pm 0.03~ \pm 0.05~$ &
          $~1.08 \pm 0.02~ \pm 0.04~$ &
	  $0.791 \pm 0.021 \pm 0.037$ 
\\
0.22    & $0.987 \pm 0.023 \pm 0.034$ &
          $0.537 \pm 0.017 \pm 0.026$ &
	  $0.317 \pm 0.013 \pm 0.024$
\\
0.28    & $0.467 \pm 0.015 \pm 0.017$ &
          $0.204 \pm 0.010 \pm 0.015$ &
	  $0.069 \pm 0.006 \pm 0.005$
\\ 
0.33    & $0.178 \pm 0.009 \pm 0.024$ &
          $0.068 \pm 0.006 \pm 0.021$ &
	  $0.008 \pm 0.002 \pm 0.007$ 
\\ 
\hline
\end{tabular}
\end{center}
\normalsize
 
Table IV.
$D_2(y_{cut})$ calculated in the E-scheme, 
the E0-scheme, and
the P-scheme (see text).
The data were corrected for detector effects and 
for initial state photon radiation.
The first error is statistical, and 
the second represents the experimental systematic uncertainty.
 
\end{table}
 
\begin{table}[b]
 
\begin{center}
\footnotesize
\renewcommand{\arraystretch}{0.8}
\begin{tabular}{|c|c|c|c|}  \hline
        &P0-scheme & D-scheme &G-scheme \\  
\raisebox{1.5ex}{$y_{cut}$} & $D_2(y_{cut}) \pm (stat.) \pm (exp.~sys.)$ &
            $D_2(y_{cut}) \pm (stat.) \pm (exp.~sys.)$ &
            $D_2(y_{cut}) \pm (stat.) \pm (exp.~sys.)$ \\
\hline
0.005   & $~39.78 \pm 0.46~ \pm 2.41~$ & 
 	  $101.06 \pm 0.74~ \pm 2.29~$ &
	  $~~7.67 \pm 0.20~ \pm 1.01~$
\\
0.010   & $~29.85 \pm 0.40~ \pm 0.78~$ &  
	  $~26.85 \pm 0.38~ \pm 0.34~$ &  
	  $~33.63 \pm 0.43~ \pm 0.84~$
\\
0.015   & $~20.49 \pm 0.33~ \pm 0.36~$ &
	  $~14.13 \pm 0.28~ \pm 0.40~$ &  
	  $~31.71 \pm 0.41~ \pm 1.01~$
\\
0.02    & $~14.52 \pm 0.28~ \pm 0.23~$ &
	  $~~9.00 \pm 0.22~ \pm 0.44~$ &  
	  $~20.46 \pm 0.33~ \pm 0.55~$
\\
0.03    & $~10.65 \pm 0.17~ \pm 0.37~$ &  
	  $~~6.02 \pm 0.13~ \pm 0.17~$ &
	  $~11.71 \pm 0.18~ \pm 0.20~$  
\\
0.05    & $~~6.36 \pm 0.09~ \pm 0.19~$ &
	  $~~3.30 \pm 0.07~ \pm 0.11~$ &
	  $~~5.55 \pm 0.09~ \pm 0.12~$
\\
0.08    & $~~3.21 \pm 0.05~ \pm 0.12~$ &
	  $~~1.66 \pm 0.04~ \pm 0.07~$ &  
	  $~~3.20 \pm 0.05~ \pm 0.06~$
\\
0.12    & $~~1.64 \pm 0.03~ \pm 0.07~$ &
	  $~0.831 \pm 0.024 \pm 0.038$ &  
	  $~~1.92 \pm 0.04~ \pm 0.05~$
\\
0.17    & $~0.944 \pm 0.023 \pm 0.057$ &
     	  $~0.406 \pm 0.015 \pm 0.033$ &
	  $~~1.25 \pm 0.03~ \pm 0.03~$  
 
\\
0.22    & $~0.433 \pm 0.015 \pm 0.038$ &
	  $~0.173 \pm 0.010 \pm 0.011$ &
	  $~0.768 \pm 0.020 \pm 0.027$  
\\
0.28    & $~0.169 \pm 0.009 \pm 0.015$ &
	  $~0.084 \pm 0.006 \pm 0.013$ &
	  $~0.409 \pm 0.014 \pm 0.019$  
\\
0.33    & $~0.034 \pm 0.004 \pm 0.008$ &
	  $~0.027 \pm 0.004 \pm 0.048$ &  
	  $~0.111 \pm 0.007 \pm 0.018$
\\ 
\hline
\end{tabular}
\end{center}
\normalsize
 
Table V.
$D_2(y_{cut})$ calculated in the P0-scheme, 
the D-scheme, and
the G-scheme (see text).
The data were corrected for detector effects and 
for initial state photon radiation.
The first error is statistical, and 
the second represents the experimental systematic uncertainty.
 
\end{table}
\newpage
\normalsize
\begin{table}[t]
\begin{center}
\footnotesize
\renewcommand{\arraystretch}{0.8}
\begin{tabular}{|c|c||c|c|}  \hline
$\chi$~(deg.)  & $EEC(rad^{-1})$  $\pm (stat.)$  $\pm (exp.~sys.)$ &
$\chi$~(deg.)  & $EEC(rad^{-1})$  $\pm (stat.)$  $\pm (exp.~sys.)$\\ 
\hline
$~~0.0 - 3.6~~$ &  ~2.265  $\pm  0.006~ $  $\pm  0.055~$ &
$~90.0 - 93.6~$ &  0.0761  $\pm  0.0009 $  $\pm  0.0013$ \\
$~~3.6 - 7.2~~$ &  ~1.316  $\pm  0.006~ $  $\pm  0.032~$ &
$~93.6 - 97.2~$ &  0.0764  $\pm  0.0009 $  $\pm  0.0025$ \\
$~~7.2 - 10.8~$ &  ~0.874  $\pm  0.004~ $  $\pm  0.020~$ &
$~97.2 - 100.8$ &  0.0777  $\pm  0.0009 $  $\pm  0.0023$ \\
$~10.8 - 14.4~$ &  ~0.598  $\pm  0.003~ $  $\pm  0.019~$ &
$100.8 - 104.4$ &  0.0809  $\pm  0.0012 $  $\pm  0.0016$ \\
$~14.4 - 18.0~$ &  ~0.425  $\pm  0.002~ $  $\pm  0.011~$ &
$104.4 - 108.0$ &  0.0834  $\pm  0.0010 $  $\pm  0.0024$ \\
$~18.0 - 21.6~$ &  ~0.310  $\pm  0.002~ $  $\pm  0.014~$ &
$108.0 - 111.6$ &  0.0874  $\pm  0.0010 $  $\pm  0.0022$ \\
$~21.6 - 25.2~$ &  ~0.241  $\pm  0.001~ $  $\pm  0.005~$ &
$111.6 - 115.2$ &  0.0931  $\pm  0.0013 $  $\pm  0.0015$ \\
$~25.2 - 28.8~$ &  ~0.199  $\pm  0.001~ $  $\pm  0.005~$ &
$115.2 - 118.8$ &  0.0968  $\pm  0.0012 $  $\pm  0.0038$ \\
$~28.8 - 32.4~$ &  ~0.168  $\pm  0.001~ $  $\pm  0.006~$ &
$118.8 - 122.4$ &  0.1030  $\pm  0.0012 $  $\pm  0.0070$ \\
$~32.4 - 36.0~$ &  ~0.146  $\pm  0.001~ $  $\pm  0.005~$ &
$122.4 - 126.0$ &  ~0.111  $\pm  0.001~ $  $\pm  0.002~$ \\
$~36.0 - 39.6~$ &  ~0.128  $\pm  0.001~ $  $\pm  0.004~$ &
$126.0 - 129.6$ &  ~0.121  $\pm  0.001~ $  $\pm  0.007~$ \\
$~39.6 - 43.2~$ &  ~0.118  $\pm  0.001~ $  $\pm  0.003~$ &
$129.6 - 133.2$ &  ~0.136  $\pm  0.002~ $  $\pm  0.003~$ \\
$~43.2 - 46.8~$ &  0.1099  $\pm  0.0008 $  $\pm  0.0026$ &
$133.2 - 136.8$ &  ~0.151  $\pm  0.002~ $  $\pm  0.004~$ \\
$~46.8 - 50.4~$ &  0.1014  $\pm  0.0009 $  $\pm  0.0031$ &
$136.8 - 140.4$ &  ~0.170  $\pm  0.002~ $  $\pm  0.005~$ \\
$~50.4 - 54.0~$ &  0.0935  $\pm  0.0008 $  $\pm  0.0027$ &
$140.4 - 144.0$ &  ~0.193  $\pm  0.002~ $  $\pm  0.006~$ \\
$~54.0 - 57.6~$ &  0.0901  $\pm  0.0009 $  $\pm  0.0021$ &
$144.0 - 147.2$ &  ~0.225  $\pm  0.002~ $  $\pm  0.008~$ \\
$~57.6 - 61.2~$ &  0.0867  $\pm  0.0008 $  $\pm  0.0023$ &
$147.2 - 151.2$ &  ~0.265  $\pm  0.002~ $  $\pm  0.007~$ \\
$~61.2 - 64.8~$ &  0.0827  $\pm  0.0009 $  $\pm  0.0023$ &
$151.2 - 154.8$ &  ~0.320  $\pm  0.003~ $  $\pm  0.008~$ \\
$~64.8 - 68.4~$ &  0.0802  $\pm  0.0010 $  $\pm  0.0018$ &
$154.8 - 158.4$ &  ~0.390  $\pm  0.003~ $  $\pm  0.013~$ \\
$~68.4 - 72.0~$ &  0.0764  $\pm  0.0009 $  $\pm  0.0031$ &
$158.4 - 162.0$ &  ~0.491  $\pm  0.003~ $  $\pm  0.017~$ \\
$~72.0 - 75.6~$ &  0.0770  $\pm  0.0010 $  $\pm  0.0010$ &
$162.0 - 165.6$ &  ~0.636  $\pm  0.004~ $  $\pm  0.012~$ \\
$~75.6 - 79.2~$ &  0.0752  $\pm  0.0008 $  $\pm  0.0031$ &
$165.6 - 169.2$ &  ~0.847  $\pm  0.006~ $  $\pm  0.007~$ \\
$~79.2 - 82.8~$ &  0.0736  $\pm  0.0008 $  $\pm  0.0013$ &
$169.2 - 172.8$ &  ~1.098  $\pm  0.005~ $  $\pm  0.009~$ \\
$~82.8 - 86.4~$ &  0.0751  $\pm  0.0010 $  $\pm  0.0015$ &
$172.8 - 176.4$ &  ~1.276  $\pm  0.007~ $  $\pm  0.044~$ \\
$~86.4 - 90.0~$ &  0.0744  $\pm  0.0010 $  $\pm  0.0014$ &
$176.4 - 180.0$ &  ~0.764  $\pm  0.007~ $  $\pm  0.050~$ \\
\hline
\end{tabular}
\end{center}
\normalsize
 
Table VI.
The $EEC$ (see text).
The data were corrected for detector effects and 
for initial state photon radiation.
The first error is statistical, and 
the second represents the experimental systematic uncertainty.
 
\end{table}
\renewcommand{\arraystretch}{1.0}
\newpage
\normalsize
\begin{table}[t]
\begin{center}
\footnotesize
\renewcommand{\arraystretch}{0.8}
\begin{tabular}{|c|c||c|c|}  \hline
$\chi$~(deg.) &$AEEC(rad^{-1})$  $\pm (stat.)$  $\pm (exp.~sys.)$  &
$\chi$~(deg.) &$JCEF(rad^{-1})$  $\pm (stat.)$  $\pm (exp.~sys.)$  
\\ \hline
$~~0.0 - 3.6~~$ &                                       &
$~90.0 - 93.6~$ &  0.0274  $\pm  0.0016$  $\pm  0.0010$ \\
$~~3.6 - 7.2~~$ &                                       & 
$~93.6 - 97.2~$ &  0.0403  $\pm  0.0020$  $\pm  0.0012$ \\
$~~7.2 - 10.8~$ &  ~0.224  $\pm  0.010~$  $\pm  0.002~$ &
$~97.2 - 100.8$ &  0.0442  $\pm  0.0026$  $\pm  0.0010$ \\
$~10.8 - 14.4~$ &  ~0.249  $\pm  0.009~$  $\pm  0.005~$ &
$100.8 - 104.4$ &  0.0523  $\pm  0.0029$  $\pm  0.0023$ \\
$~14.4 - 18.0~$ &  ~0.211  $\pm  0.006~$  $\pm  0.005~$ &
$104.4 - 108.0$ &  0.0566  $\pm  0.0029$  $\pm  0.0024$ \\
$~18.0 - 21.6~$ &  ~0.181  $\pm  0.004~$  $\pm  0.005~$ &
$108.0 - 111.6$ &  0.0613  $\pm  0.0034$  $\pm  0.0026$ \\
$~21.6 - 25.2~$ &  ~0.148  $\pm  0.004~$  $\pm  0.006~$ &
$111.6 - 115.2$ &  0.0725  $\pm  0.0039$  $\pm  0.0017$ \\
$~25.2 - 28.8~$ &  ~0.121  $\pm  0.003~$  $\pm  0.004~$ &
$115.2 - 118.8$ &  0.0832  $\pm  0.0055$  $\pm  0.0046$ \\
$~28.8 - 32.4~$ &  0.0972  $\pm  0.0024$  $\pm  0.0029$ &
$118.8 - 122.4$ &  0.0858  $\pm  0.0051$  $\pm  0.0016$ \\
$~32.4 - 36.0~$ &  0.0785  $\pm  0.0022$  $\pm  0.0062$ &
$122.4 - 126.0$ &  0.0944  $\pm  0.0043$  $\pm  0.0024$ \\
$~36.0 - 39.6~$ &  0.0645  $\pm  0.0017$  $\pm  0.0024$ &
$126.0 - 129.6$ &  0.1051  $\pm  0.0061$  $\pm  0.0055$ \\
$~39.6 - 43.2~$ &  0.0513  $\pm  0.0020$  $\pm  0.0026$ &
$129.6 - 133.2$ &  ~0.114  $\pm  0.005~$  $\pm  0.002~$ \\
$~43.2 - 46.8~$ &  0.0413  $\pm  0.0015$  $\pm  0.0027$ &
$133.2 - 136.8$ &  ~0.131  $\pm  0.005~$  $\pm  0.005~$ \\
$~46.8 - 50.4~$ &  0.0346  $\pm  0.0016$  $\pm  0.0021$ &
$136.8 - 140.4$ &  ~0.148  $\pm  0.005~$  $\pm  0.006~$ \\
$~50.4 - 54.0~$ &  0.0275  $\pm  0.0013$  $\pm  0.0060$ &
$140.4 - 144.0$ &  ~0.169  $\pm  0.007~$  $\pm  0.004~$ \\
$~54.0 - 57.6~$ &  0.0213  $\pm  0.0010$  $\pm  0.0024$ &
$144.0 - 147.2$ &  ~0.188  $\pm  0.007~$  $\pm  0.005~$ \\
$~57.6 - 61.2~$ &  0.0163  $\pm  0.0008$  $\pm  0.0073$ &
$147.2 - 151.2$ &  ~0.228  $\pm  0.008~$  $\pm  0.009~$ \\
$~61.2 - 64.8~$ &  0.0141  $\pm  0.0007$  $\pm  0.0026$ &
$151.2 - 154.8$ &  ~0.275  $\pm  0.009~$  $\pm  0.010~$ \\
$~64.8 - 68.4~$ &  0.0129  $\pm  0.0010$  $\pm  0.0008$ &
$154.8 - 158.4$ &  ~0.329  $\pm  0.011~$  $\pm  0.013~$ \\
$~68.4 - 72.0~$ &  0.0110  $\pm  0.0007$  $\pm  0.0025$ &
$158.4 - 162.0$ &  ~0.414  $\pm  0.011~$  $\pm  0.019~$ \\
$~72.0 - 75.6~$ &  0.0064  $\pm  0.0005$  $\pm  0.0017$ &
$162.0 - 165.6$ &  ~0.551  $\pm  0.012~$  $\pm  0.013~$ \\
$~75.6 - 79.2~$ &  0.0058  $\pm  0.0006$  $\pm  0.0029$ &
$165.6 - 169.2$ &  ~0.751  $\pm  0.021~$  $\pm  0.021~$ \\
$~79.2 - 82.8~$ &  0.0041  $\pm  0.0004$  $\pm  0.0020$ &
$169.2 - 172.8$ &  ~1.095  $\pm  0.024~$  $\pm  0.019~$ \\
$~82.8 - 86.4~$ &  0.0012  $\pm  0.0002$  $\pm  0.0038$ &
$172.8 - 176.4$ &  ~1.639  $\pm  0.032~$  $\pm  0.034~$ \\
$~86.4 - 90.0~$ &  0.0017  $\pm  0.0008$  $\pm  0.0016$ &
$176.4 - 180.0$ &  ~1.530  $\pm  0.039~$  $\pm  0.049~$ \\
\hline
\end{tabular}
\end{center}
\normalsize
 
Table VII.
The $AEEC$ and $JCEF$ (see text).
The data were corrected for detector effects and 
for initial state photon radiation.
The first error is statistical, and 
the second represents the experimental systematic uncertainty.
 
\end{table}
\renewcommand{\arraystretch}{1.0}
 
\newpage
\clearpage
%
%
\begin{table}[t]
\footnotesize
\vspace{0.5cm}
\begin{center}
\begin{tabular}{|c|c|r|c|c|c|c|c|}  \hline 
           &           &               &        &
\multicolumn{4}{c|}{uncertainties} \\ \cline{5-8}
\raisebox{1.5ex}{observable} & \raisebox{1.5ex}{fit range} &
\raisebox{1.5ex}{$f$-range~~~~} & \raisebox{1.5ex}{$\asz$} & stat. & 
exp. sys. & had. & scale \\
\hline \hline
$\tau$   & $0.06-0.32$ & $2\times 10^{-4} - 4$ & 0.1245 &
           $\pm 0.0008$ & $\pm 0.0017$ & $\pm 0.0026$ & $\pm 0.0201$\\ \hline
$\rho$   & $0.04-0.32$ & $1.5\times 10^{-3} - 4$ & 0.1273 &
           $\pm 0.0008$ & $\pm 0.0020$ & $\pm 0.0005$ & $\pm 0.0096$\\ \hline
$B_T$    & $0.12-0.32$ & $5.7\times 10^{-3} - 4$ & 0.1272 &
           $\pm 0.0008$ & $\pm 0.0020$ & $\pm 0.0033$ & $\pm 0.0220$\\ \hline
$B_W$    & $0.06-0.26$ & $2\times 10^{-3} - 4$ & 0.1196 &
           $\pm 0.0008$ & $\pm 0.0026$ & $\pm 0.0024$ & $\pm 0.0072$\\ \hline
$O$      & $0.08-0.32$ & $2\times 10^{-1} - 4$ & 0.1343 &
           $\pm 0.0013$ & $\pm 0.0015$ & $\pm 0.0087$ & $\pm 0.0082$\\ \hline
$C$      & $0.24-0.76$ & $4\times 10^{-4} -  4$ & 0.1233 &
           $\pm 0.0009$ & $\pm 0.0019$ & $\pm 0.0032$ & $\pm 0.0186$\\ \hline
\hline
$D_2(\mbox{E})$    & $0.08-0.28$ & $5\times 10^{-5} - 4$ & 0.1273 &
           $\pm 0.0006$ & $\pm 0.0016$ & $\pm 0.0022$ & $\pm 0.0217$\\ \hline
$D_2(\mbox{E0})$ & $0.05-0.28$ & $1.2\times 10^{-2} - 4$ & 0.1175 &
           $\pm 0.0007$ & $\pm 0.0027$ & $\pm 0.0010$ & $\pm 0.0083$\\ \hline
$D_2(\mbox{P})$    & $0.05-0.22$ & $5.5\times 10^{-3} - 4$ & 0.1207 &
           $\pm 0.0008$ & $\pm 0.0033$ & $\pm 0.0025$ & $\pm 0.0053$\\ \hline
$D_2(\mbox{P0})$ & $0.05-0.28$ & $1.2\times 10^{-2} - 4$ & 0.1190 & 
           $\pm 0.0009$ & $\pm 0.0031$ & $\pm 0.0020$ & $\pm 0.0057$\\ \hline
$D_2(\mbox{D})$    & $0.03-0.22$ & $1.7\times 10^{-3} - 4$ & 0.1245 &
           $\pm 0.0011$ & $\pm 0.0032$ & $\pm 0.0007$ & $\pm 0.0077$\\ \hline
$D_2(\mbox{G})$    & $0.12-0.28$ & $4\times 10^{-3} - 4$ & 0.1191 &
           $\pm 0.0008$ & $\pm 0.0014$ & $\pm 0.0029$ & $\pm 0.0043$\\ \hline
\hline
$EEC$  & $~36.0^{\circ}-154.8^{\circ}$ & $3.5\times 10^{-3} -4$ & 0.1222 &
           $\pm 0.0008$ & $\pm 0.0030$ & $\pm 0.0021$ & $\pm 0.0121$\\ \hline
$AEEC$ & $~18.0^{\circ}-68.4^{\circ}~$ & $9\times 10^{-2} - 4$ & 0.1121 &
           $\pm 0.0012$ & $\pm 0.0032$ & $\pm 0.0017$ & $\pm 0.0031$\\ \hline
\hline
$JCEF$ & $100.8^{\circ}-158.4^{\circ}$ & $5\times 10^{-3} - 4$ & 0.1185 &
           $\pm 0.0007$ & $\pm 0.0027$ & $\pm 0.0008$ & $\pm 0.0045$\\ \hline
\end{tabular}
\end{center}
\normalsize
Table VIII.
Observables used in $\Oaa$ QCD fits.
For each the fit range,
the range of the renormalization scale factor considered,
central $\asz$ value, statistical and experimental systematic errors, 
and hadronization and scale uncertainties are shown. 
 
\end{table}

\newpage
\clearpage
 
%
%
\begin{table}[t]
\footnotesize
\begin{center}
\begin{tabular}{|c||c|c|c|}  \hline 
observable &  $\alpha_s(M_Z^2)$ & exp. error & theoretical uncertainty
\\ \hline \hline
$\tau$   & 0.1245 & $\pm 0.0019$ & $\pm 0.0203$ \\ \hline
$\rho$   & 0.1273 & $\pm 0.0022$ & $\pm 0.0096$ \\ \hline
$B_T$    & 0.1272 & $\pm 0.0022$ & $\pm 0.0222$ \\ \hline
$B_W$    & 0.1196 & $\pm 0.0027$ & $\pm 0.0076$ \\ \hline
$O$      & 0.1343 & $\pm 0.0020$ & $\pm 0.0120$ \\ \hline
$C$      & 0.1233 & $\pm 0.0021$ & $\pm 0.0189$ \\ \hline \hline
$D_2(\mbox{E})$ & 0.1273 & $\pm 0.0017$ & $\pm 0.0218$ \\ \hline
$D_2(\mbox{E0})$& 0.1175 & $\pm 0.0028$ & $\pm 0.0084$ \\ \hline
$D_2(\mbox{P})$ & 0.1207 & $\pm 0.0034$ & $\pm 0.0059$ \\ \hline
$D_2(\mbox{P0})$& 0.1190 & $\pm 0.0032$ & $\pm 0.0060$ \\ \hline
$D_2(\mbox{D})$ & 0.1245 & $\pm 0.0034$ & $\pm 0.0077$ \\ \hline
$D_2(\mbox{G})$ & 0.1191 & $\pm 0.0016$ & $\pm 0.0052$ \\ \hline \hline
$EEC$  & 0.1222 & $\pm 0.0031$ & $\pm 0.0123$ \\ \hline
$AEEC$ & 0.1121 & $\pm 0.0034$ & $\pm 0.0035$ \\ \hline \hline
$JCEF$ & 0.1185 & $\pm 0.0028$ & $\pm 0.0046$ \\ \hline
\end{tabular}
\end{center}
\normalsize
Table IX.
The $\alpha_s(M_Z^2)$ values derived from $\Oaa$ QCD fits.
 
\end{table}

\newpage
%
%
 
\begin{table}[t]
\footnotesize
\begin{center}
\begin{tabular}{|c|c|c|c|c|c|}  \hline 
           &           &
                   ln$R$ matching &
                   mod. ln$R$ matching &
                   $R$ matching &
                   mod. $R$ matching 
\\  \cline{3-6}  
observable & fit range & 
		   $\asz \pm \Delta \alpha_s$ & 
                   $\asz \pm \Delta \alpha_s$ & 
                   $\asz \pm \Delta \alpha_s$ & 
                   $\asz \pm \Delta \alpha_s$ 
\\ 
           &           &
                   $f$-range &
                   $f$-range &
                   $f$-range &
                   $f$-range 
\\ \hline \hline
$\tau$   & $0.06-0.32$ & $0.1196\pm 0.0089$ & 
                         $0.1203\pm 0.0089$ & 
                         $0.1226\pm 0.0110$ & 
                         $0.1187\pm 0.0091$ 
\\
         &             & $2.7\times 10^{-3}- 4$ &
	                 $2.7\times 10^{-3}- 4$ &
			 $1.9\times 10^{-3}- 4$ &
			 $2.3\times 10^{-3}- 4$  
\\ \hline
$\rho$   & $0.04-0.32$ & $0.1151\pm 0.0039$ & 
                         $0.1162\pm 0.0047$ & 
                         $0.1178\pm 0.0061$ & 
                         $0.1146\pm 0.0044$ 
\\
         &             & $1.1\times 10^{-2}- 4$ &
	                 $1.1\times 10^{-2}- 4$ &
			 $4.9\times 10^{-3}- 4$ &
			 $1.0\times 10^{-2}- 4$  
\\ \hline
$B_T$    & $0.12-0.32$ & $0.1175\pm 0.0030$ & 
                         $0.1211\pm 0.0015$ & 
                         $-$                & 
                         $0.1177\pm 0.0017$ 
\\
         &             & $6.7\times 10^{-2}- 4$ &
	                 $3.0\times 10^{-1}- 4$ &
			                           &
			 $3.6\times 10^{-2}- 4$  
\\ \hline
$B_W$    & $0.06-0.26$ & $0.1083\pm 0.0016$ & 
                         $0.1095\pm 0.0003$ & 
                         $-$               & 
                         $0.1107\pm 0.0034$ 
\\
         &             & $8.2\times 10^{-2}- 4$ &
	                 $1.9\times 10^{-1}- 4$ &
			                            &
			 $4.9\times 10^{-2}- 4$  
\\ \hline \hline
$D_2(\mbox{D})$ & $0.03-0.22$ & $0.1312\pm 0.0060$ & 
                         $0.1313\pm 0.0059$ & 
                         $0.1251\pm 0.0053$ &
                         \mbox{N/A}
\\
         &             & $1.5\times 10^{-1}- 4$ &
	                 $1.6\times 10^{-1}- 4$ &
			 $7.0\times 10^{-2}- 4$ &
			 
\\ \hline \hline
 
$EEC$   & $90.0^{\circ}-154.8^{\circ}$ & \mbox{N/A} & 
                                       \mbox{N/A} & 
                                       $0.1239\pm 0.0049$ & 
                                       $0.1336\pm 0.0028$ 
\\
         &             & &
	                 &
			 $6.1\times 10^{-2}- 4$ &
			 $2.7\times 10^{-1}- 4$  
\\ \hline
\end{tabular}
\end{center}
\normalsize
Table X.
Observables used in resummed$+\Oaa$ fits.
For each the fit range, the range of the renormalization scale 
factor considered,
the central $\asz$ value, and scale uncertainty ($\Delta \as$) are given.
Results are shown separately for each of the four matching schemes considered.
Acceptable fits to the data could not be obtained for $B_T$ and $B_W$
with the \Rm \ scheme.   
 
\end{table}
 
\newpage
\clearpage
%
%
\begin{table}[t]
\footnotesize
\begin{center}
\begin{tabular}{|c||c|c|c|}  \hline 
observable &  $\alpha_s(M_Z^2)$ & exp. error & theoretical uncertainty
\\ \hline \hline
$\tau$          & 0.1180 & $\pm 0.0018$ & $\pm 0.0115$ \\ \hline
$\rho$   	& 0.1163 & $\pm 0.0020$ & $\pm 0.0064$ \\ \hline
$B_T$    	& 0.1160 & $\pm 0.0020$ & $\pm 0.0048$ \\ \hline 
$B_W$    	& 0.1074 & $\pm 0.0025$ & $\pm 0.0042$ \\ \hline \hline
$D_2(\mbox{D})$ & 0.1297 & $\pm 0.0035$ & $\pm 0.0073$ \\ \hline \hline
$EEC$    	& 0.1279 & $\pm 0.0032$ & $\pm 0.0069$ \\ \hline
\end{tabular}
\end{center}
\normalsize
Table XI.
The $\alpha_s(M_Z^2)$ values derived from resummed$+\Oaa$ QCD fits.
\end{table}
 
\clearpage
 
\newpage
%
%
\begin{table}[b]
\footnotesize
\begin{center}
\begin{tabular}{|c|c|c|c|c|c|}  \hline 
           &           &
                   ln$R$ matching &
                   mod. ln$R$ matching &
                   $R$ matching &
                   mod. $R$ matching 
\\  \cline{3-6}  
observable & fit range & 
		   $\asz \pm \Delta \alpha_s$ & 
                   $\asz \pm \Delta \alpha_s$ & 
                   $\asz \pm \Delta \alpha_s$ & 
                   $\asz \pm \Delta \alpha_s$ 
\\ 
           &           &
                   $f$-range &
                   $f$-range &
                   $f$-range &
                   $f$-range 
\\ \hline \hline
$\tau$   & $0.02-0.32$ & $0.1170\pm 0.0086$ & 
                         $0.1184\pm 0.0075$ & 
                         $-$              & 
                         $0.1191\pm 0.0045$ 
\\
         &             & $7.0\times 10^{-2} - 4$ &
	                 $1.4\times 10^{-1} - 4$ &
			                            &
			 $6.3\times 10^{-1} - 4$  
\\ \hline
$\rho$   & $0.02-0.32$ & $0.1153\pm 0.0071$ & 
                         $0.1146\pm 0.0072$ & 
                         $0.1140\pm 0.0054$ & 
                         $0.1124\pm 0.0071$ 
\\
         &             & $2.6\times 10^{-2} - 4$ &
	                 $3.4\times 10^{-2} - 4$ &
			 $2.0\times 10^{-1} - 4$ &
			 $4.0\times 10^{-2} - 4$  
\\ \hline
$B_T$    & $0.04-0.32$ & $0.1177\pm 0.0040$ & 
                         $0.1202\pm 0.0021$ & 
                         $-$                & 
                         $0.1175\pm 0.0023$ 
\\
         &             & $2.0\times 10^{-1} - 4$ &
	                 $6.7\times 10^{-2} - 4$ &
			                           &
			 $1.1\times 10^{-1} - 4$  
\\ \hline
$B_W$    & $0.04-0.26$ & $0.1078\pm 0.0024$ & 
                         $0.1089\pm 0.0014$ & 
                         $-$              & 
                         $0.1106\pm 0.0032$ 
\\
         &             & $1.4\times 10^{-1} - 4$ &
	                 $2.8\times 10^{-1} - 4$ &
			                            &
			 $5.4\times 10^{-2} - 4$  
\\ \hline \hline
$D_2(\mbox{D})$ & $0.01-0.22$ & $0.1269\pm 0.0026$ & 
                         $0.1268\pm 0.0025$ & 
                         $-$ &
                         \mbox{N/A}
\\
         &             & $1.3\times 10^{-1} - 4$ &
	                 $1.3\times 10^{-1} - 4$ &
			                            &
			 
\\ \hline \hline
 
$EEC$   & $90.0^{\circ}-162.0^{\circ}$ & \mbox{N/A} & 
                                       \mbox{N/A} & 
                                       $0.1233\pm 0.0043$ & 
                                       $0.1337\pm 0.0027$
\\
         &             & &
	                 &
			 $6.9\times 10^{-2} - 4$ &
			 $5.0\times 10^{-1} - 4$  
\\ \hline
\end{tabular}
\end{center}
\normalsize
Table XII.
Observables used in resummed$+\Oaa$ fits with the fit ranges extended 
into the two-jet region.
For each the fit range, the range of the renormalization scale 
factor considered,
the central $\asz$ value, and scale uncertainty ($\Delta \as$) are given.
Results are shown separately for each of the four matching schemes considered.
Acceptable fits to the data could not be obtained for 
$\tau$, $B_T$, $B_W$, and $D_2$(D) with the \Rm \ scheme.

\end{table}
 
\newpage
\clearpage
%
%
\begin{table}[t]
\footnotesize
\begin{center}
\begin{tabular}{|c||c|c|c|}  \hline 
observable &  $\alpha_s(M_Z^2)$ & exp. error & theoretical uncertainty
\\ \hline \hline
$\tau$          & 0.1159 & $\pm 0.0017$ & $\pm 0.0090$ \\ \hline
$\rho$          & 0.1144 & $\pm 0.0019$ & $\pm 0.0074$ \\ \hline
$B_T$           & 0.1157 & $\pm 0.0020$ & $\pm 0.0053$ \\ \hline 
$B_W$           & 0.1070 & $\pm 0.0025$ & $\pm 0.0041$ \\ \hline \hline
$D_2(\mbox{D})$ & 0.1274 & $\pm 0.0034$ & $\pm 0.0027$ \\ \hline \hline
$EEC$           & 0.1285 & $\pm 0.0032$ & $\pm 0.0068$ \\ \hline
\end{tabular}
\end{center}
\normalsize
Table XIII.
The $\alpha_s(M_Z^2)$ values derived from resummed$+\Oaa$ QCD fits
with the fit ranges extended into the two-jet region.
\end{table}
 
\clearpage
 
\newpage
 
\clearpage
\section*{Figure captions }
 
\noindent
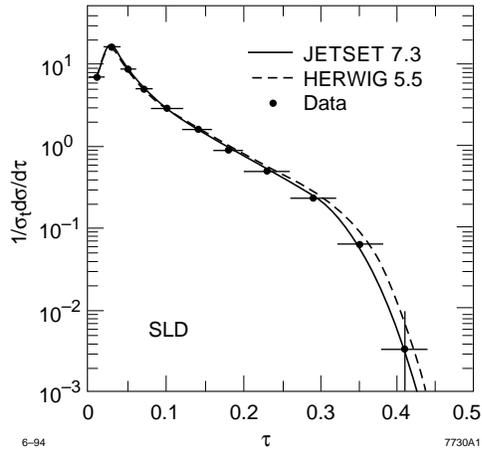
{\bf Figure 1}.	
The measured thrust distribution corrected to the hadron level.
The error bars include the statistical and experimental systematic errors
added in quadrature. 
The curves show the predictions of the QCD parton shower models 
JETSET 7.3 (solid line) and HERWIG 5.5 (dashed line).
 
\noindent
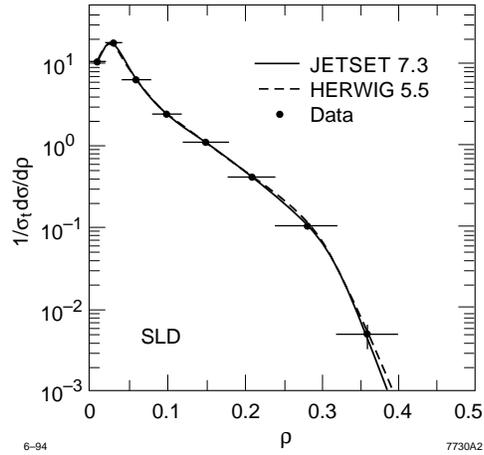
{\bf Figure 2}.
As Fig. 1 but for the heavy jet mass.
 
\noindent
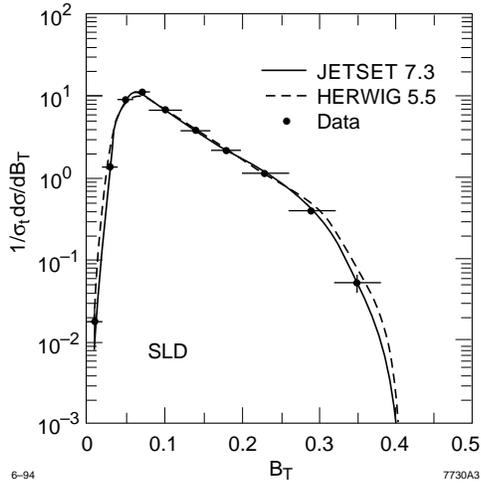
{\bf Figure 3}.
As Fig. 1 but for the total jet broadening.
 
\noindent
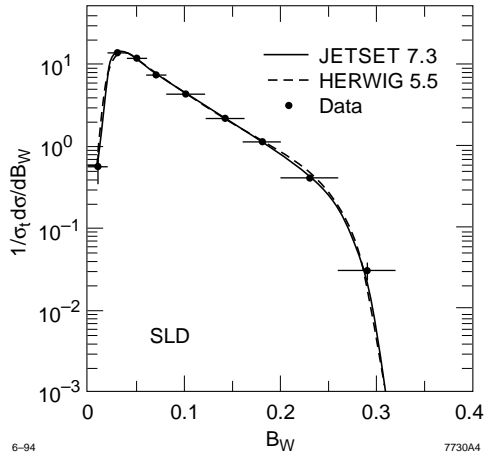
{\bf Figure 4}.
As Fig. 1 but for the wide jet broadening.
 
\noindent
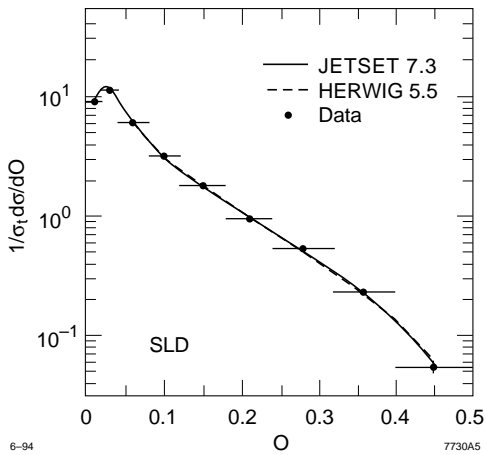
{\bf Figure 5}.
As Fig. 1 but for the oblateness.
 
\noindent
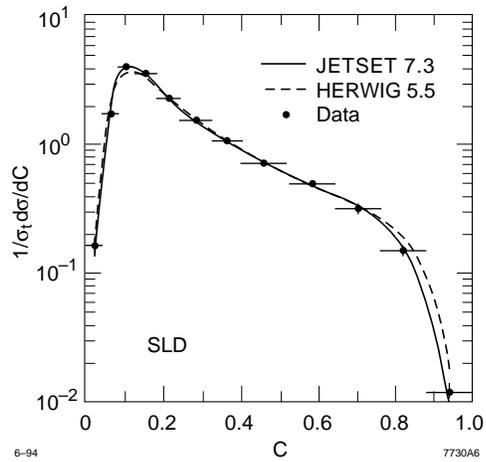
{\bf Figure 6}.
As Fig. 1 but for the C-parameter.
 
\noindent
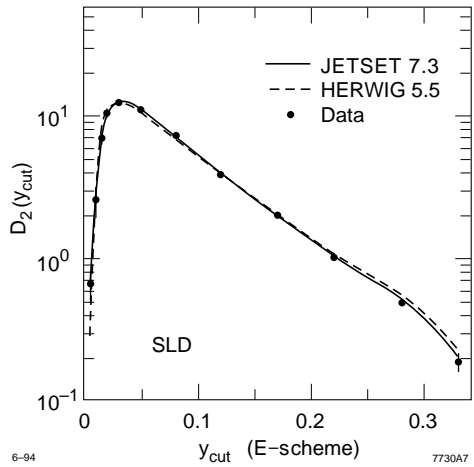
{\bf Figure 7}.
As Fig. 1 but for the differential 2-jet rate with the E-scheme.
 
\noindent
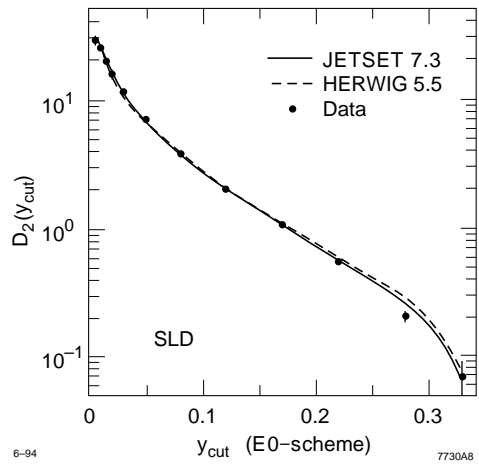
{\bf Figure 8}.
As Fig. 1 but for the differential 2-jet rate with the E0-scheme.
 
\noindent
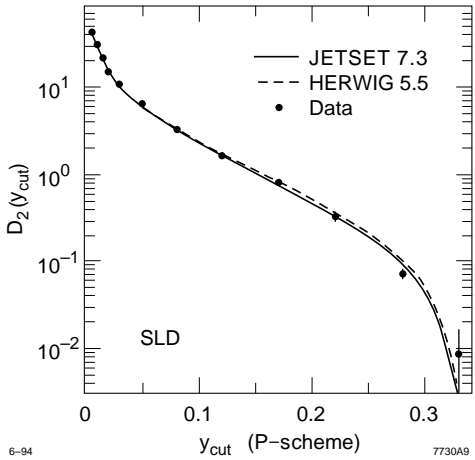
{\bf Figure 9}.
As Fig. 1 but for the differential 2-jet rate with the P-scheme.
 
\noindent
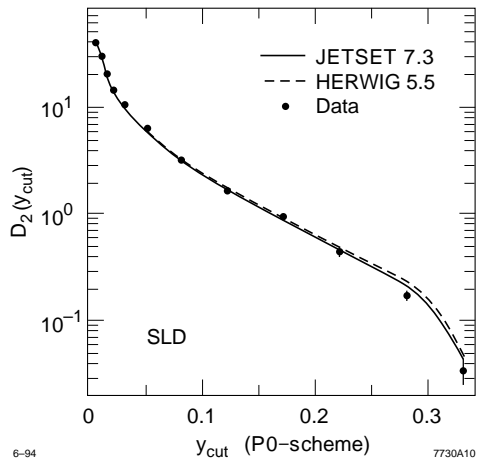
{\bf Figure 10}.
As Fig. 1 but for the differential 2-jet rate with the P0-scheme.
 
\noindent
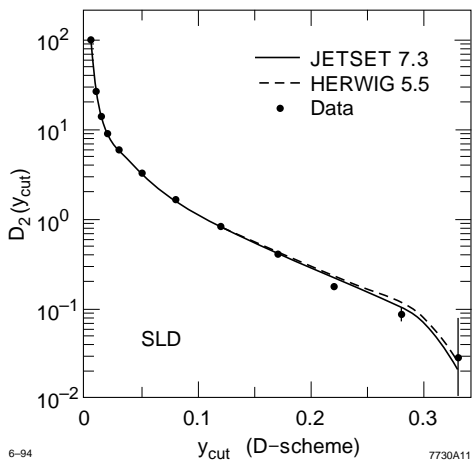
{\bf Figure 11}.
As Fig. 1 but for the differential 2-jet rate with the D-scheme. 
 
\noindent
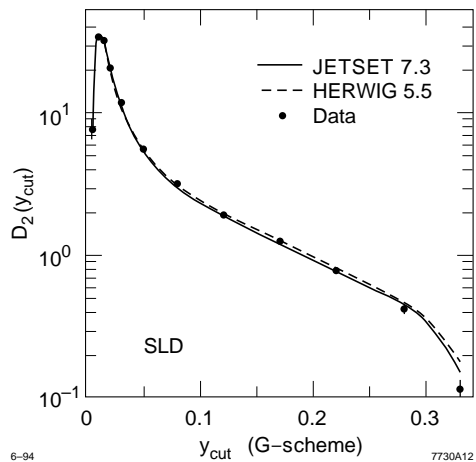
{\bf Figure 12}.
As Fig. 1 but for the  differential 2-jet rate with the G-scheme.
 
\noindent
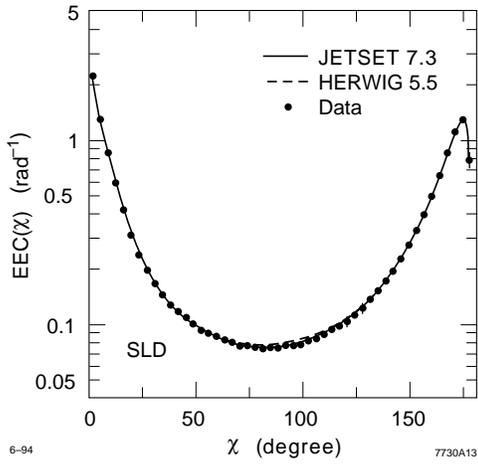
{\bf Figure 13}.
As Fig. 1 but for the energy-energy correlation $EEC$.
 
\noindent
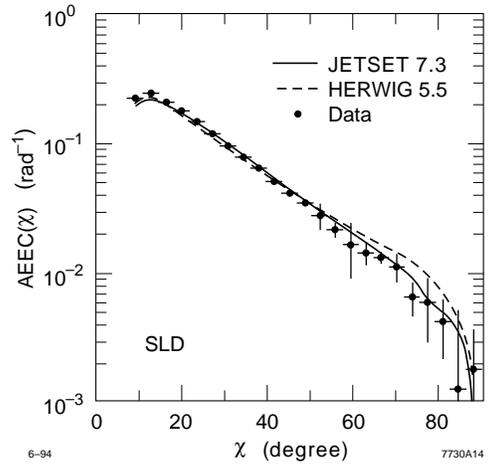
{\bf Figure 14}.
As Fig. 1 but for the asymmetry of the energy-energy correlation $AEEC$.
 
\noindent
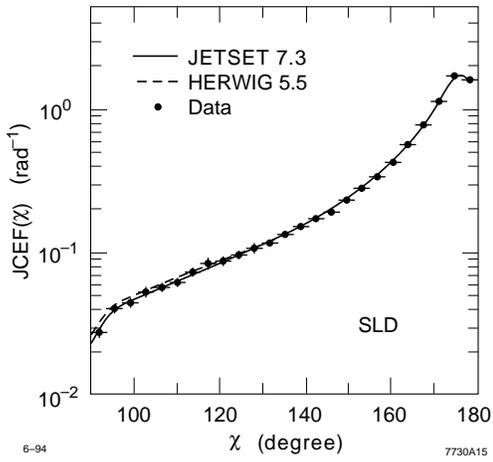
{\bf Figure 15}.
As Fig. 1 but for the jet cone energy fraction $JCEF$.
 
\noindent
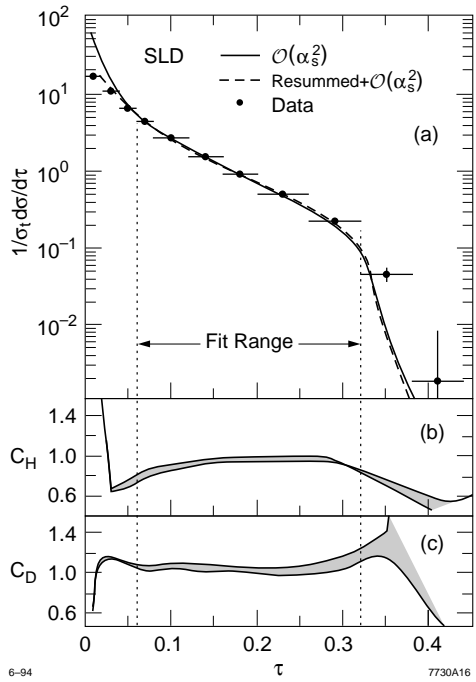
{\bf Figure 16}.	
(a) The measured thrust distribution corrected to the parton level.
The error bars include the statistical and experimental systematic errors
added in quadrature. 
The curves show the predictions of the $O(\alpha_s^2)$ 
calculations (solid line) and
the resummed$+\Oaa$ calculations with {\it modified lnR-matching} (dashed line).
The renormalization scale factor was fixed to 1.
Sizes of the (b) hadronization correction and 
(c) detector correction
factors; 
the widths of the bands indicate the systematic uncertainties.
 
\noindent
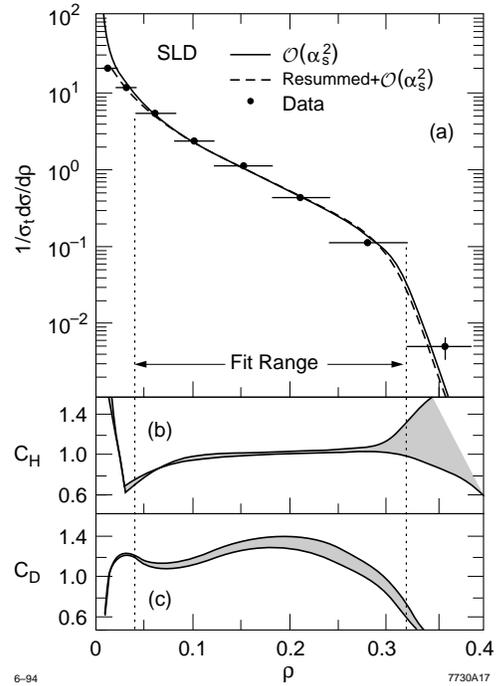
{\bf Figure 17}.
As Fig. 16 but for the heavy jet mass.
 
\noindent
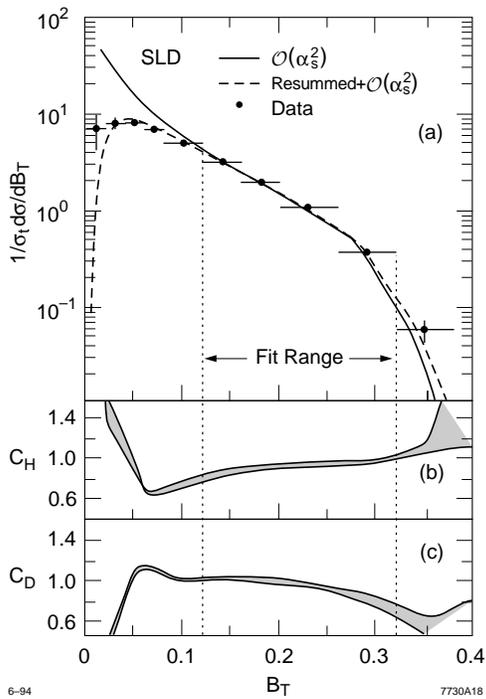
{\bf Figure 18}.
As Fig. 16 but for the total jet broadening.
 
\noindent
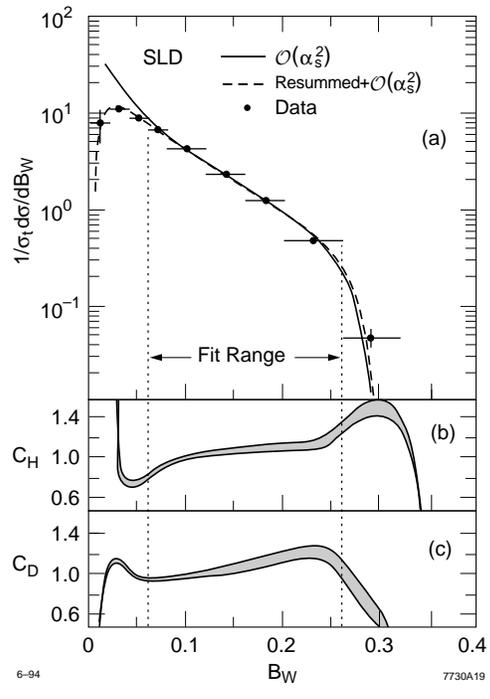
{\bf Figure 19}.
As Fig. 16 but for the wide jet broadening.
 
\noindent
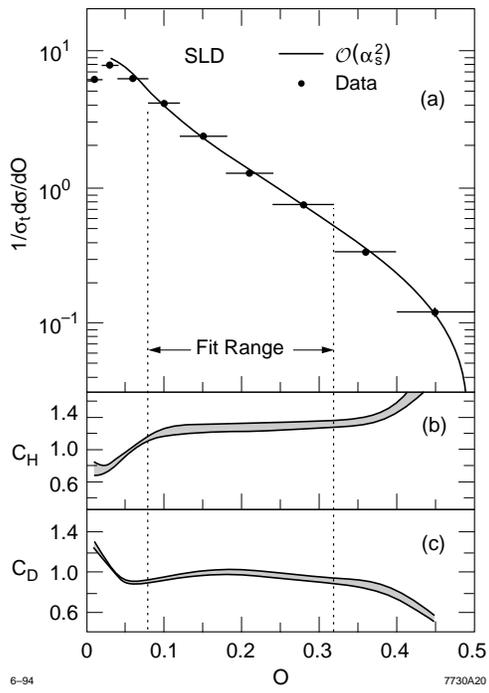
{\bf Figure 20}.
As Fig. 16 but for the oblateness.
 
\noindent
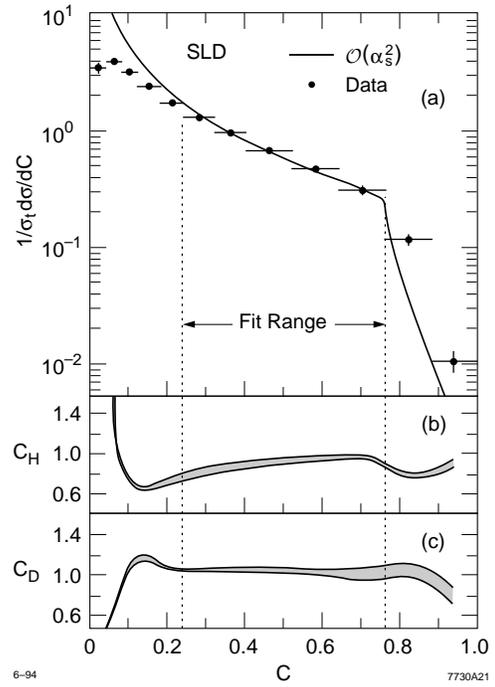
{\bf Figure 21}.
As Fig. 16 but for the C-parameter.
 
\noindent
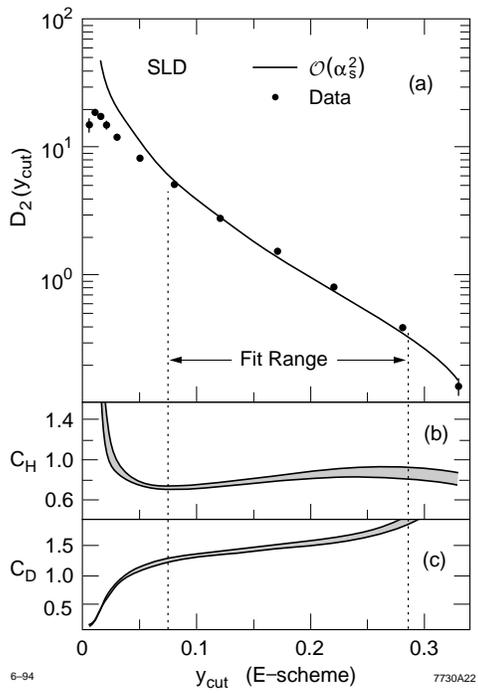
{\bf Figure 22}.
As Fig. 16 but for the differential 2-jet rate with the E-scheme.
 
\noindent
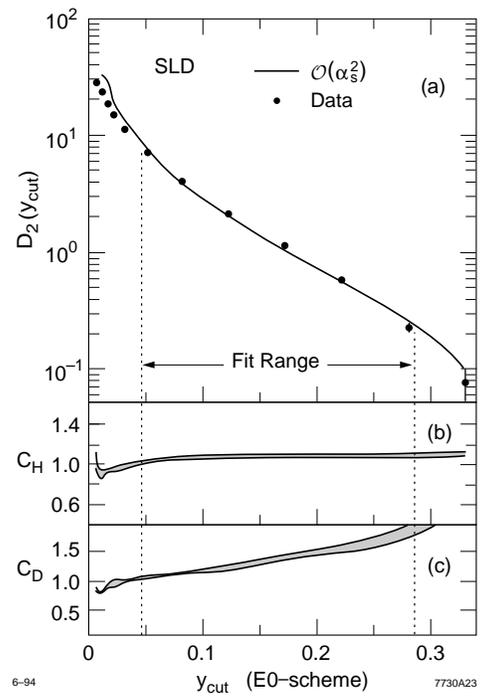
{\bf Figure 23}.
As Fig. 16 but for the differential 2-jet rate with the E0-scheme.
 
\noindent
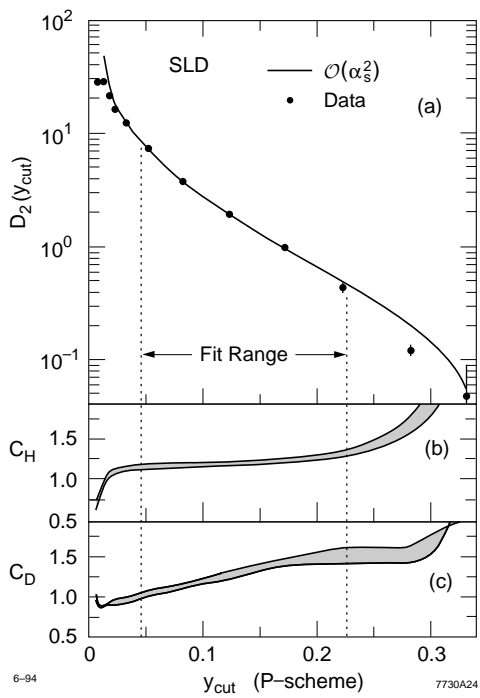
{\bf Figure 24}.
As Fig. 16 but for the differential 2-jet rate with the P-scheme.
 
\noindent
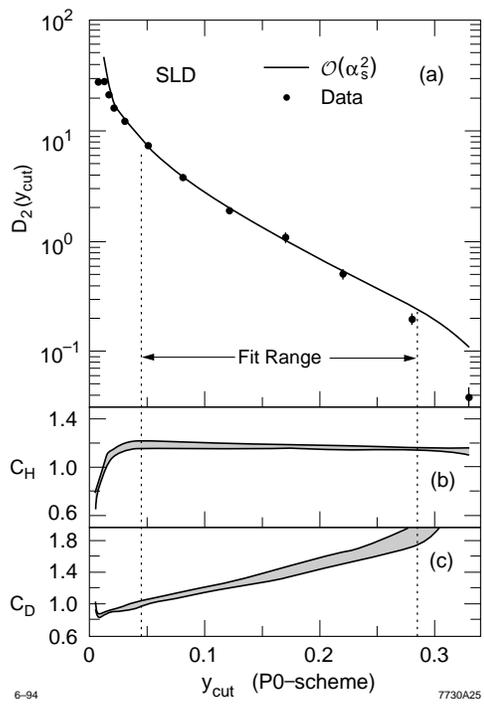
{\bf Figure 25}.
As Fig. 16 but for the differential 2-jet rate with the P0-scheme.
 
\noindent
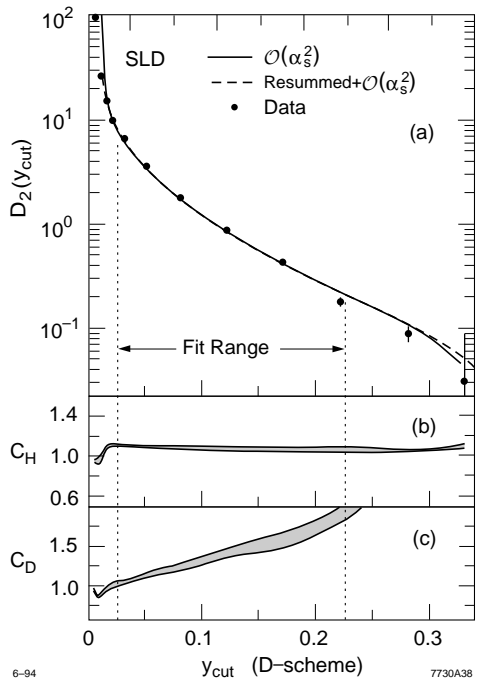
{\bf Figure 26}.
As Fig. 16 but for the differential 2-jet rate with the D-scheme. 
 
\noindent
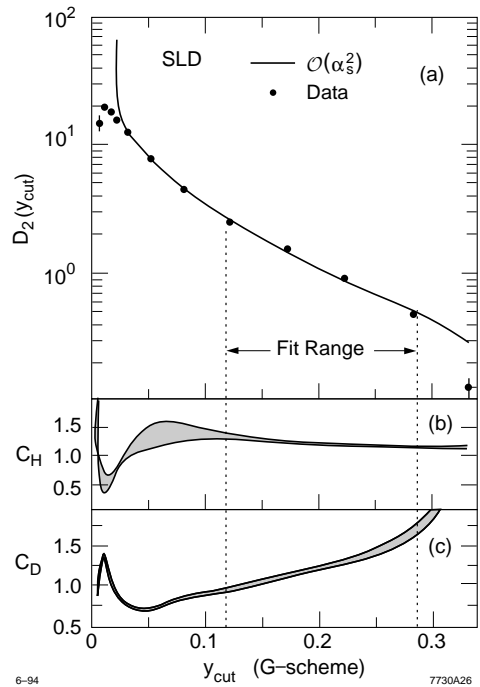
{\bf Figure 27}.
As Fig. 16 but for the  differential 2-jet rate with the G-scheme.
 
\noindent
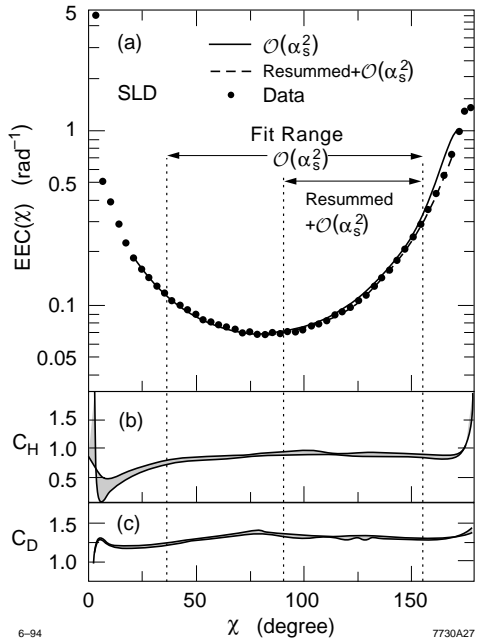
{\bf Figure 28}.
As Fig. 16 but for the energy-energy correlation $EEC$.
 
\noindent
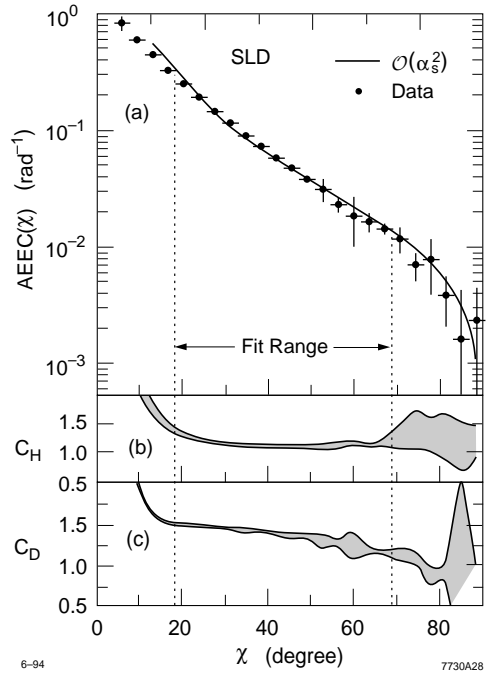
{\bf Figure 29}.
As Fig. 16 but for the asymmetry of the energy-energy correlation $AEEC$.
 
\noindent
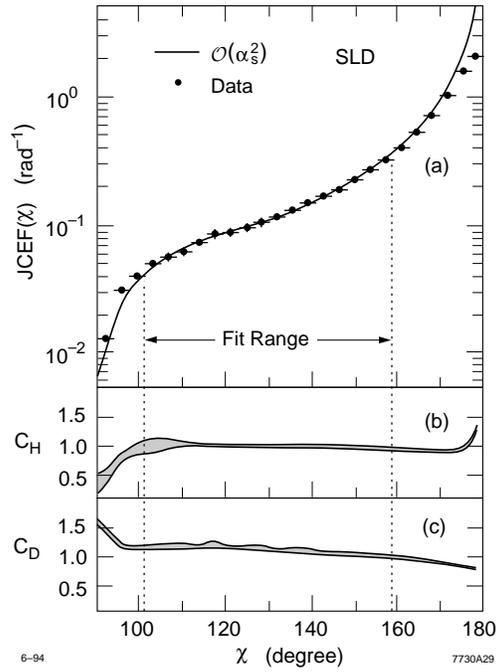
{\bf Figure 30}.
As Fig. 16 but for the jet cone energy fraction $JCEF$.
 
\noindent
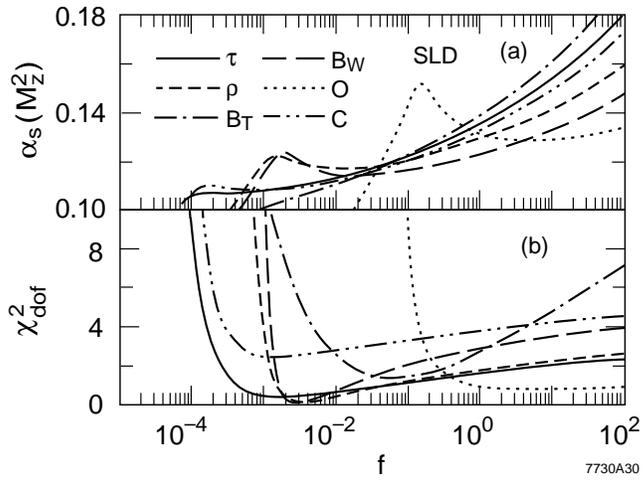
{\bf Figure 31}.	
(a) $\asz$ and (b) $\chi^2_{dof}$ 
from the $\Oaa$ fits to the event shapes as a function of
renormalization scale factor $f$ (see text).
 
\noindent
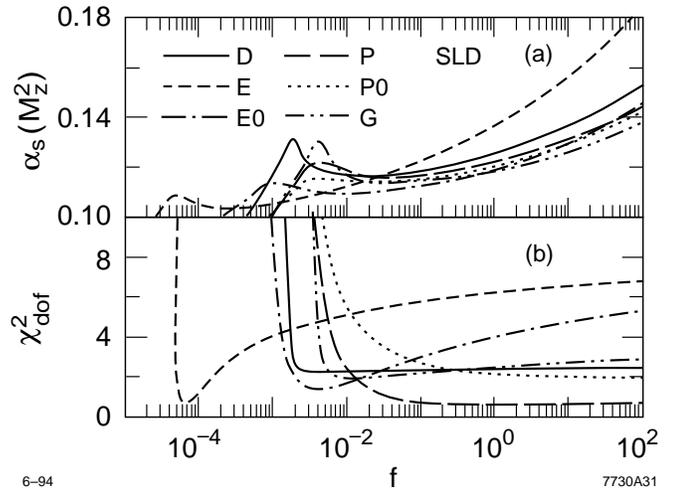
{\bf Figure 32}.	
(a) $\asz$ and (b) $\chi^2_{dof}$ 
from the $\Oaa$ fits to the jet rates as a function of
renormalization scale factor $f$ (see text).
 
\noindent
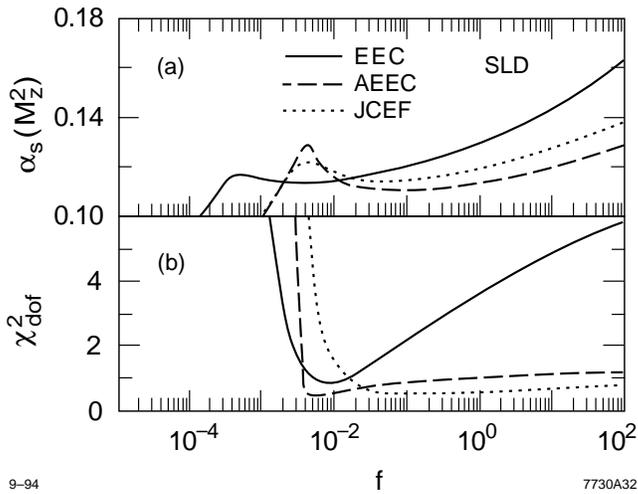
{\bf Figure 33}.	
(a) $\asz$ and (b) $\chi^2_{dof}$ 
from the $\Oaa$ fits to the particle correlations and angular
energy flow as a function of
renormalization scale factor $f$ (see text).
 
\noindent
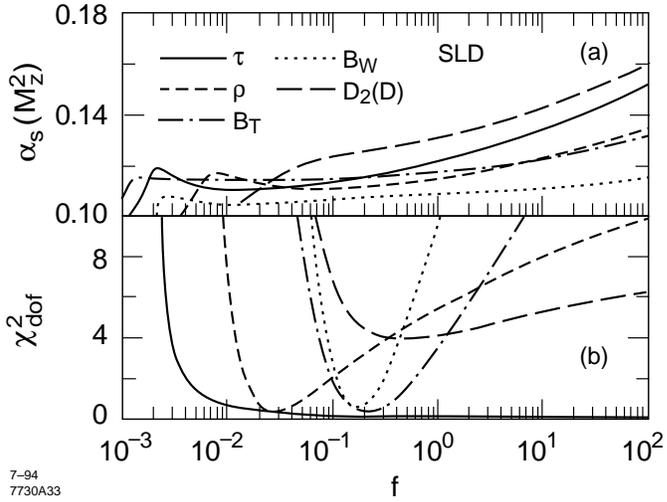
{\bf Figure 34}.	
(a) $\asz$ and (b) $\chi^2_{dof}$ 
from the resummed$+\Oaa$ fits with $lnR$-$matching$ 
as a function of renormalization scale factor $f$ (see text).
 
\noindent
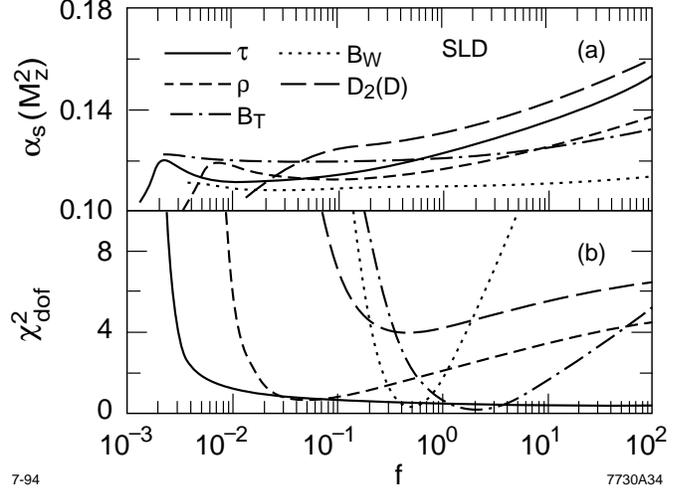
{\bf Figure 35}.	
(a) $\asz$ and (b) $\chi^2_{dof}$ 
from the resummed$+\Oaa$ fits with $modified~lnR$-$matching$
as a function of renormalization scale factor $f$ (see text).
 
\noindent
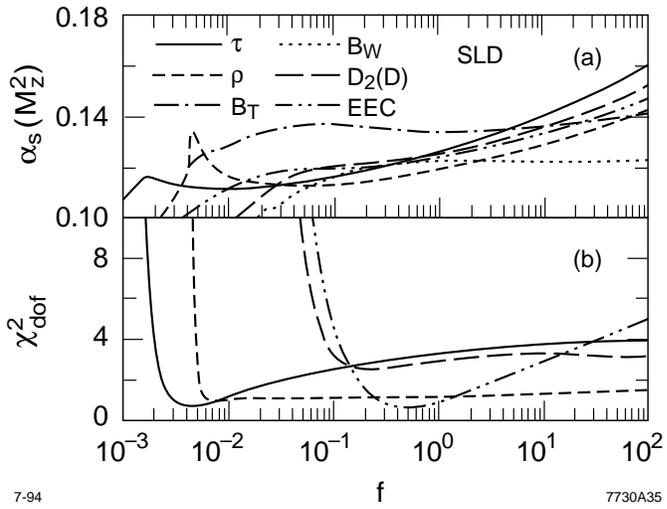
{\bf Figure 36}.	
(a) $\asz$ and (b) $\chi^2_{dof}$ 
from the resummed$+\Oaa$ fits with $R$-$matching$
as a function of renormalization scale factor $f$ (see text).
The $\chi^2_{dof}$ values for $B_T$ and $B_W$ are larger than 10
for all $f$.
 
\noindent
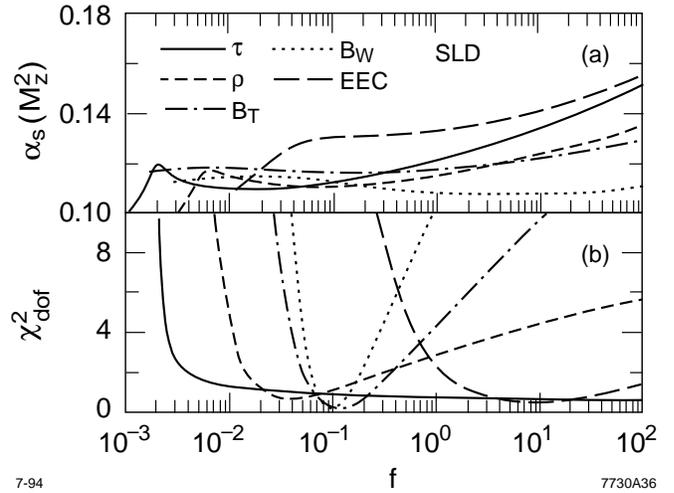
{\bf Figure 37}.	
(a) $\asz$ and (b) $\chi^2_{dof}$ 
from the resummed$+\Oaa$ fits with $modified~R$-$matching$
as a function of renormalization scale factor $f$ (see text).
 
\noindent
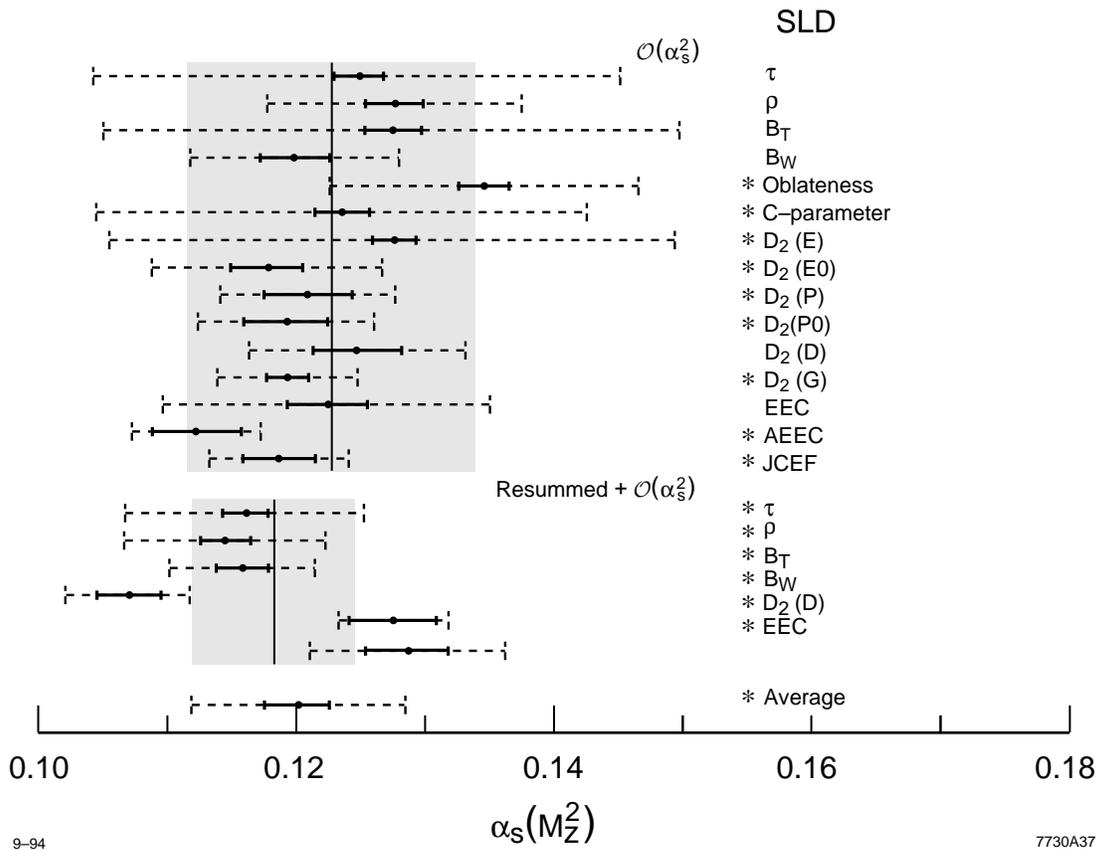
{\bf Figure 38}.	Compilation of final values of $\asz$.
For each observable 
the solid bar denotes the experimental error,
while the dashed  bar shows the total uncertainty
comprising the experimental error and theoretical uncertainty
in quadrature.
Shown separately for the $\Oaa$ results and resummed$+\Oaa$ results are
a vertical line and a shaded region representing
the average $\asz$ value and uncertainty, respectively, in each case.
Also shown is the final average of six resummed$+\Oaa$ and nine $\Oaa$
results indicated by stars.

\end{document}